\newif\ifreport
\reporttrue

\ifreport
\documentclass[a4paper]{article}
\usepackage{layaureo}
\else
\documentclass[conference]{IEEEtran}
\fi
\usepackage{lipsum}
\usepackage{amsmath}
\usepackage[T1]{fontenc}
\usepackage[]{graphicx}
\usepackage{pgfplots}
\usepackage{pgfplotstable}
\usepackage{dcolumn}
\usepackage{url}
\usepackage{booktabs}
\usepackage{xspace}
\usepackage{stmaryrd}
\usepackage{wasysym}
\usepackage{array}
\usepackage{mathtools,amssymb}
\usetikzlibrary{decorations.markings}
\usepackage{tikz}
\pgfdeclarelayer{edgelayer}
\pgfdeclarelayer{nodelayer}
\usetikzlibrary{arrows, shapes.geometric}
\usepackage{tkz-euclide}
\usepackage[font=footnotesize]{subcaption}

\newcommand{\sol}{SOLOIST\xspace}
\newcommand{\ct}{CLTLB($\mathcal{D}$)\xspace}
\newcommand{\qf}{QF-EUFIDL\xspace}
\usepackage{paralist}
\usepackage{tabularx}
\usepackage{mathptmx}
\usepackage[stable]{footmisc}

\newcommand{\plotHeight}{6.5cm}
\newcommand{\plotWidth}{0.7\textwidth}
\newcommand\blfootnote[1]{%
  \begingroup
  \renewcommand\thefootnote{}\footnote{#1}%
  \addtocounter{footnote}{-1}%
  \endgroup
}

\title{Offline Trace Checking of Quantitative Properties of Service-Based Applications}

\ifreport
\author{
Domenico Bianculli\\
SnT Centre - University of Luxembourg, Luxembourg\\
domenico.bianculli@uni.lu\\
\and
Carlo Ghezzi, 
Sr\dj{}an Krsti\'c,
 Pierluigi San Pietro\\
DEEP-SE group - DEIB - Politecnico di Milano, Italy\\
\{carlo.ghezzi,srdan.krstic,pierluigi.sanpietro\}@polimi.it\\
}
\else
\author{\IEEEauthorblockN{Domenico Bianculli}
\IEEEauthorblockA{SnT Centre - University of Luxembourg, Luxembourg\\
Email: domenico.bianculli@uni.lu}
\and
\IEEEauthorblockN{Carlo Ghezzi \qquad Sr\dj{}an Krsti\'c  \qquad Pierluigi San Pietro}
\IEEEauthorblockA{DEEP-SE group - DEIB - Politecnico di Milano, Italy\\
Email:\{carlo.ghezzi,srdan.krstic,pierluigi.sanpietro\}@polimi.it}
}
\fi

\begin{document}

\maketitle

\ifreport
\else
\blfootnote{This work has been partially supported by  the National Research Fund, Luxembourg (FNR/P10/03).}
\fi

\begin{abstract}
Service-based applications are often developed as compositions of
partner services.  A service integrator needs precise methods to
specify the quality attributes expected by each partner service, as
well as effective techniques to verify these attributes.  In previous
work, we identified the most common specification patterns related to
provisioning service-based applications and developed an expressive
specification language (SOLOIST) that supports them.
SOLOIST is an extension of metric temporal logic with aggregate
temporal modalities that can be used to write quantitative temporal
properties.

In this paper we address the problem of performing offline checking of
service execution traces against quantitative requirements
specifications written in SOLOIST. We present a translation of SOLOIST into CLTLB($\mathcal{D}$), a
variant of linear temporal logic, and reduce the trace checking of SOLOIST to bounded
satisfiability checking of CLTLB($\mathcal{D}$), which is supported by
 ZOT, an SMT-based verification toolkit.  
We detail the results of applying the proposed offline trace
checking procedure to different types of traces, and compare its
performance with previous work.

%%% Local Variables: 
%%% mode: latex
%%% TeX-master: "soca2014"
%%% End: 

\end{abstract}

%\keywords{Trace checking, metric temporal logic, aggregate operators,
 % bounded model checking, SMT}

\section{Introduction}
\label{sec:introduction}

% Services (The new (old) approach)
Service-based applications (SBAs) are one of the main approaches
followed nowadays to develop modern enterprise information systems,
adopting the paradigm of service-oriented
computing~\cite{Josuttis:2007:SPA:1207973}.  SBAs are usually defined
as service compositions, created by orchestrating several existing
services, possibly provided by third-parties, by means of dedicated
languages such as BPEL.  Developing and operating an SBA involves many
stakeholders: service end-users, the developers and providers of
services used in the SBA, as well as the service integrators that
realize the composite services. However, service integrators have the
ultimate responsibility for maintaining an adequate level of quality
attributes (e.g., in terms of functional correctness and QoS, quality
of service) of the composite services they provide, independently of
(but at the same time, based on) the guarantees and the service-level
agreements offered by the providers of the services they compose.
This can be achieved in a systematic and formal way by developing a
specification language that can capture useful properties of SBAs and
by providing means for verifying SBAs against properties written in
such a specification language.

Several verification techniques have been developed and
tailored~\cite{bozkurt2012:testing--verifi,salaun2010:analysis-and-ve,canfora2008:service-oriente,
  2007:test-and-analys} for the domain of SBAs, to assist service
integrators in verification activities both at design time (e.g.,
testing, model checking) and run time (e.g., monitoring).  In the case
of formal approaches, the verification techniques adopt a temporal
logic (such as LTL, CTL) as the specification language of the
properties of interest. In the domain of composite SBAs, these
properties express constraints on the interactions of the composite
service with its partner services.
% spec patterns + soloist
In a previous work some of the authors developed
\sol (\emph{SpecificatiOn Language fOr servIce compoSitions 
inTeractions})~\cite{bianculli13:_tale_solois}
a metric temporal logic with new, additional
temporal modalities that can express properties of SBAs in terms 
of bounds on some aggregated values, calculated over a certain time window.
These modalities have been defined based on an extensive field study~\cite{bgps:icse2012} 
of the requirements specifications in the context of service-based applications, 
and they are tailored to express the most common requirements occurring in practice. 
The study --- performed in collaboration with an industrial partner ---				
analyzed more than 900 requirements specifications, extracted both from research
papers and industrial data, and led to the identification of a new
class of specification patterns,
specific to the domain of service provisioning. Examples of these
patterns are those characterizing the \emph{average response time} of
a service invocation and 
the \emph{count/average/maximum number of event occurrences} in a
given time window.

%Trace checking
In this paper we focus on the problem of performing \emph{offline
   checking} of execution traces against requirements
specifications written in \sol.  Trace checking (also called
\emph{trace validation}~\cite{mrad13:_babel} or \emph{history
  checking}~\cite{Felder:1994:VRS:201024.201034}) is a procedure for
evaluating a formal specification over a log of recorded events
produced by a system, i.e., over a temporal evolution of the system.
We assume that a trace is finite and composed by the events
corresponding to the interactions of a composite service with its
external services (e.g., invoking external service operations or
receiving service requests). 
Traces can be produced 
by a proper monitoring/logging infrastructure, and made
available at the end of the execution to perform offline trace
checking.

% contribution
The main contribution of the paper is an offline trace checking procedure for
\sol properties exploiting a translation into
\ct~\cite{bersani12:_const_ltl_satis_check_autom}, an
extension of PLTLB (Propositional Linear Temporal Logic with both past
and future modalities) augmented with atomic formulae built over a
constraint system $\mathcal{D}$. We chose \ct as the
target of our translation since it supports the definition of
arithmetical constraints over a set of integer variables (also called
counters); as we will detail in Sect.~\ref{sec:basic-translation},
these counters allow a compact and easy-to-verify 
translation.  We express the problem of trace checking of \sol
properties in terms of bounded satisfiability checking (BSC) of
\ct and rely on the BSC procedure for metric
temporal logic~\cite{pradella2013:bounded-satisfi} implemented in 
ZOT.  We focus on requirements containing quantitative
properties involving aggregate operations on events occurring in a
given time window, like the average response time of a certain
operation provided by a partner service.

%Previous work
% LTL translation
In the original definition of \sol~\cite{bianculli13:_tale_solois} we
showed how, under certain assumptions, the language can be translated
into LTL, guaranteeing its decidability based
on well-known results in temporal logic. However, this translation was
only a proof of concept and was not meant to guarantee efficiency if one 
would use LTL-based verification procedures.
%QF-EUFIDL translation
In previous work~\cite{bbgks-fase2014} we introduced a trace checking
procedure, based on another encoding of \sol properties into formulae of
QF-EUFIDL, the theory of quantifier-free integer difference logic with
uninterpreted function and predicate symbols. This encoding was
tailored for sparse traces, i.e., traces in which the number of time
instants when events occur is very low with respect to the length of
the trace. In contrast, the new encoding proposed in this paper
supports a much more efficient checking of dense traces.
In this paper we also compare the two approaches on
traces of different degrees of sparseness.

\ifreport
The rest of the paper is organized as
follows. 
\else
The rest of the paper is organized as
follows.\footnote{A more complete version of the paper can be found at \url{}}
\fi
Section~\ref{sec:preliminaries}
provides background information on \sol and  \ct.
%, and introduces a running example.
 We present the translation of \sol into
\ct in Section~\ref{sec:basic-translation}. 
In section~\ref{sec:implementation} we discuss some implementation
details. 
Section~\ref{sec:evaluation} reports on the evaluation performed to
assess the scalability of our approach, also in comparison with
previous work.
Section~\ref{sec:related-work} surveys related work, and 
Sect.~\ref{sec:concl-future-work} concludes the paper, giving some directions for future work.

%%% Local Variables: 
%%% mode: latex
%%% TeX-master: "soca2014"
%%% End: 

\section{Preliminaries}
\label{sec:preliminaries}

\subsection{\sol in a Nutshell}
\label{sec:soloist-nutshell}

\ifreport
Previous work~\cite{bgps:icse2012} reports the results of
a field study on the the specification patterns used to express
requirements of SBAs. After analyzing more than 900 requirements
specifications extracted both from research papers and industrial data, the authors of
the study identified a set of specification patterns specific to
service provisioning (in addition to the well-known ones like those defined
in~\cite{konrad2005:real-time-speci,dwyer1998:property-specif}). The
service provisioning patterns refer to: 
\begin{inparaenum}[S1)]
  \item average response time;
  \item counting the number of events;
  \item average number of events;
  \item maximum number of events;
  \item absolute time;
  \item unbounded elapsed time;
  \item data-awareness.
\end{inparaenum}
On the basis of these findings, a new specification language, called 
\sol (\emph{SpecificatiOn Language fOr servIce compoSitions
inTeractions}), was introduced in ~\cite{bianculli13:_tale_solois}. 
\sol was in fact designed with the goal of supporting the common  specification patterns found for service
provisioning; it is a propositional 
metric temporal logic with new temporal modalities that support
aggregate operations on events occurring in a given time window.
\else
In this section we provide a brief overview of \sol; a
rationale of the language and a detailed explanation of its
semantics are in~\cite{bianculli13:_tale_solois}.
\fi

The syntax of \sol is defined by the following grammar:\\
$\phi \Coloneqq p \mid  \neg \phi \mid \phi \land
\phi \mid \phi \mathsf{U}_I \phi \mid \phi
\mathsf{S}_I \phi \mid \mathfrak{C}^{K}_{\bowtie n}(\phi) \mid
\mathfrak{U}^{K,h}_{\bowtie n}(\phi) \mid \mathfrak{M}^{K,h}_{\bowtie
  n}(\phi) \mid \mathfrak{D}_{\bowtie n}^K
  (\phi,\phi)$\\
where $p \in \Pi$, with $\Pi$ being a finite set of atoms; $I$  is a nonempty interval over $\mathbb{N}$;  $n,K,h$
range over $\mathbb{N}$; $\bowtie \  \in \{<,\leq,\geq,>,=\}$. 
The arguments $\phi$ of modalities $\mathfrak{C}, \mathfrak{U}, \mathfrak{M},
\mathfrak{D}$ are restricted to atoms in $\Pi$.
 Moreover, the two arguments in the $\mathfrak{D}$ modality are
 required to be different atoms.

The $\mathsf{U}_I$ and $\mathsf{S}_I$ modalities are, respectively,
the metric ``\emph{Until}'' and ``\emph{Since}'' operators. Additional
temporal modalities can be derived using the usual conventions; for
example ``\emph{Always}'' is defined as $\mathsf{G}_I \phi \equiv \neg
(\top \mathsf{U}_I \neg \phi)$ and ``\emph{Eventually in the Past}''
as $\mathsf{P}_I \phi \equiv \top \mathsf{S}_I \phi$, where $\top$
means ``true''.  The remaining modalities are called \emph{aggregate} modalities. 
The $\mathfrak{C}^{K}_{\bowtie n}(\phi)$ modality
states a bound (represented by $\bowtie n$) on the number of occurrences of an event $\phi$
in the previous $K$ time instants; 
\ifreport
it expresses pattern S2.
\else
it is used to count the number of events in a given time window.
\fi
The $\mathfrak{U}^{K,h}_{\bowtie n}(\phi)$ (respectively,
$\mathfrak{M}^{K,h}_{\bowtie n}(\phi)$) modality expresses a bound on
the average (respectively, maximum) number of occurrences of an event
$\phi$, aggregated over the set of right-aligned adjacent
non-overlapping subintervals within a time window $K$; 
\ifreport
it corresponds to pattern S3 (respectively, S4).
\else
as in ``the average/maximum number of events per hour in the last ten hours''.  
\fi
A subtle difference in the
semantics of the $\mathfrak{U}$ and $\mathfrak{M}$ modalities is that
$\mathfrak{M}$ considers events in the (possibly empty) tail interval,
i.e., the leftmost observation subinterval whose length is less than $h$,
while the $\mathfrak{U}$ modality ignores them. 
The $\mathfrak{D}^{K}_{\bowtie n}(\phi,\psi)$ modality expresses a bound
on the average time elapsed between a pair of specific adjacent events
$\phi$ and $\psi$ occurring in the previous $K$ time
instants; 
\ifreport
 it can be used to express pattern S1.
\else
it is used to express the concept of average response
time (of a service), with $\phi$ and $\psi$ representing, respectively the start and
end events, of a (synchronous) service invocation\footnote{As detailed in~\cite{bianculli13:_tale_solois}, we
  assume that two subsequent occurrences of the $\phi$ or $\psi$ events
  may not happen. This models the behavior of synchronous service
  invocations.}.
\fi

\begin{figure}[tb]
\begin{center}
\ifreport
\begin{small}
\else
\begin{tiny}
\fi
\begin{tabular}{>{$}p{25mm}<{$}c>{$}l<{$}}
   %%%%%%%%%%%%%%%%%%%%%%%%%%%%%%%%%%%%%%%%%%%%%%%%%%%%%%%%%%%%
   %p(t1,...,tn)
   (w, i)  \models
   p &{if{f}}& p \in \sigma_i\\
   %%%%%%%%%%%%%%%%%%%%%%%%%%%%%%%%%%%%%%%%%%%%%%%%%%%%%%%%%%%%
   %not phi
   (w, i)  \models \neg \phi
   &{if{f}}&  (w, i) \not
   \models \phi \\
   %%%%%%%%%%%%%%%%%%%%%%%%%%%%%%%%%%%%%%%%%%%%%%%%%%%%%%%%%%%%
   % and
   (w, i)  \models \phi \land \psi
   &{if{f}}& (w, i) 
   \models \phi \land (w, i)
   \models \psi\\
   %%%%%%%%%%%%%%%%%%%%%%%%%%%%%%%%%%%%%%%%%%%%%%%%%%%%%%%%%%%%
   % since
   (w, i)  \models \phi \mathsf{S}_I
   \psi &{if{f}}& \text{for some } j < i, \tau_i - \tau_j \in
   I,  (w,j)  \models \psi
    \text{ and for all } k, j < k < i,  (w, k)  
   \models \phi\\
   %%%%%%%%%%%%%%%%%%%%%%%%%%%%%%%%%%%%%%%%%%%%%%%%%%%%%%%%%%%%
   % until
   (w, i)  \models \phi \mathsf{U}_I
   \psi &{if{f}}& \text{for some } j > i, \tau_j - \tau_i \in
   I,  (w,j)  \models \psi
    \text{ and for all }k, i < k < j,  (w, k)  
   \models \phi\\
   %%%%%%%%%%%%%%%%%%%%%%%%%%%%%%%%%%%%%%%%%%%%%%%%%%%%%%%%%%%%
   % C count
   (w, i)  \models
   \mathfrak{C}_{\bowtie n}^{K}(\phi) &{if{f}}& 
   c(\tau_i - K, \tau_i,\phi) \bowtie n \text{ and } \tau_i \geq K\\
   %%%%%%%%%%%%%%%%%%%%%%%%%%%%%%%%%%%%%%%%%%%%%%%%%%%%%%%%%%%%
   % V average number of events
   (w, i)  \models
   \mathfrak{U}_{\bowtie n}^{K,h}(\phi) &{if{f}}&
   \dfrac{ c(\tau_i - \lfloor \frac{K}{h}\rfloor h, \tau_i,\phi)}{\lfloor \frac{K}{h} \rfloor} \bowtie n \text{ and } \tau_i \geq K
   % \newline \text{(*) alternative semantics} \newline
   % \dfrac{c(\tau_i - K, \tau_i, \bar{\mathcal{D}}, \bar{\tau},
   %   \sigma,\phi)}{\frac{K}{h}} \bowtie n 
   %  \newline  \text{(**) alternative semantics} \newline
   % \dfrac{c(\tau_i - K, \tau_i, \bar{\mathcal{D}}, \bar{\tau},
   %   \sigma,\phi)}{\lfloor \frac{K}{h} \rfloor} \bowtie n
    \\
   %%%%%%%%%%%%%%%%%%%%%%%%%%%%%%%%%%%%%%%%%%%%%%%%%%%%%%%%%%%%
   % M maximum number of events
   (w, i)  \models
   \mathfrak{M}_{\bowtie n}^{K,h}(\phi) &{if{f}}& 
  \max\left\{\bigcup_{m=0}^{\left \lfloor \frac{K}{h} \right \rfloor}
     \left\{ c(\mathit{lb}(m),\mathit{rb}(m),\phi) \right\} \right\}
   \bowtie n \text{ and } \tau_i \geq K\\
   %%%%%%%%%%%%%%%%%%%%%%%%%%%%%%%%%%%%%%%%%%%%%%%%%%%%%%%%%%%%
   % D average distance
   (w, i)  \models \mathfrak{D}_{\bowtie n}^K
   (\phi,\psi)  &{if{f}}&
   \dfrac{\sum_{(s,t) \in d(\phi,\psi,\tau_i,K)}
     (\tau_t-\tau_s)}{|d(\phi,\psi,\tau_i,K)|}
   \bowtie n  \text{ and } \tau_i \geq K \text{ and }d(\phi,\psi,\tau_i,K)\neq\emptyset \\[5mm]
   %%%%%%%%%%%%%%%%%%%%%%%%%%%%%%%%%%%%%%%%%%%%%%%%%%%%%%%%%%%%
   \multicolumn{3}{l}{where
$c(\tau_a,\tau_b,\phi) = | \left \{ s \mid
\tau_a < \tau_s \leq \tau_b  \text{ and } (w, s) \models \phi \right
\} |$, $\mathit{lb}(m)=\max\{\tau_i - K,  \tau_i - (m+1) h \} $,
}\\
\ifreport
\multicolumn{3}{l}{$\mathit{rb}(m)=\tau_i - mh$, and $d(\phi,\psi,\tau_i, K) = \{(s,t) \mid \tau_i - K <
  \tau_s \leq \tau_i \text{ and } (w, s) \models \phi, $}\\
\multicolumn{3}{l}{ $ t= \min\{ u \mid \tau_s < \tau_u \leq \tau_i, (w,u)  \models \psi\} \}$}\\
\else
\multicolumn{3}{l}{$\mathit{rb}(m)=\tau_i - mh$, and $d(\phi,\psi,\tau_i, K) = \{(s,t) \mid \tau_i - K <
  \tau_s \leq \tau_i \text{ and } (w, s) \models \phi, $ $ t= \min\{ u \mid \tau_s < \tau_u \leq
    \tau_i, (w,u)  \models \psi\} \}$}
\fi
% \\
%  \multicolumn{3}{l}{$ t= \min\{ u \mid \tau_s < \tau_u \leq
%     \tau_i, (w,u)  \models \psi\} \}$}
  
\end{tabular}

%%% Local Variables: 
%%% mode: latex
%%% TeX-master: "../issta2014"
%%% End: 

\ifreport
\end{small}
\else
\end{tiny}
\fi
\end{center}
\caption{Formal semantics of \sol}
\label{fig:semantics}\end{figure}

 The formal semantics of \sol is trace-based, i.e.,
 defined
on timed $\omega$-words over
$2^{\Pi} \times \mathbb{N}$.
A timed sequence $\tau = \tau_0 \tau_1 \ldots $ is an infinite sequence of values $\tau_i \in \mathbb{N}$
satisfying $\tau_i<\tau_{i+1}$, for all $i\geq 0$, i.e., the sequence
increases strictly monotonically.
A timed $\omega$-word over alphabet $2^{\Pi}$ is a pair $(\sigma, \tau)$ where $\sigma= \sigma_0 \sigma_1 \ldots$ is an infinite word
over $2^{\Pi}$ and $\tau$ is a timed sequence. A timed language over $2^{\Pi}$ is a set of timed words over the same alphabet.  
%PL: removed the following definition since it is never used in the paper:
% A \sol formula is said to characterize a timed language  over $2^{\Pi}$ such
% that its words  satisfy the formula. 
Notice that there is a distinction between the integer position $i$ in the timed $\omega$-word and the corresponding integer timestamp $\tau_i$.
Figure~\ref{fig:semantics} defines the satisfiability relation $(w, i) \models \phi$ 
for every timed $\omega$-word $w$, every position $i\ge 0$ and for every \sol formula $\phi$.
%For the sake of simplicity, hereafter we express the $\mathfrak{U}$
%modality in terms of the  $\mathfrak{C}$ one, based on this
%definition:  $\mathfrak{U}_{\bowtie n}^{K,h} (\phi) \equiv
%\mathfrak{C}_{\bowtie n \cdot \lfloor \frac{K}{h} \rfloor}^{\lfloor \frac{K}{h} \rfloor \cdot h} (\phi) 
%$, which can be derived from the semantics in Fig.~\ref{fig:semantics}.

We remark that the version of \sol presented here is a restriction of the original one 
in~\cite{bianculli13:_tale_solois}. To simplify the presentation in the
next sections, we dropped first-order quantification on finite domains (which was introduced to support 
\ifreport
 data-awareness, i.e., pattern S7
\else
data-carrying events
\fi
) and  limited the
argument of the $\mathfrak{D}$ modality to only one pair of
events.
These restrictions are only syntactic sugar and we refer to~\cite{bianculli13:_tale_solois} for the details of the transformations that provide support for them.

\ifreport
\subsection{Example}
\label{sec:running}

In this section we show how to express common quantitative properties
of an SBA by means of \sol.  As an example, we consider an SBA realized as a service
composition described in BPEL, depicted in Fig.~\ref{fig:process}
using the (visually intuitive) notation introduced in~\cite{bbggs:iet07}.

The process \emph{ATMFrontEnd} starts when the \textit{receive}
activity \texttt{logOn} processes a message from the
\textit{SessionManager} service. This starts a customer session: the
process verifies whether the customer holds a valid account at the
bank, by invoking the \texttt{checkAccess} operation of the
\textit{BankAccount} service. If the latter identifies the customer, a
loop is started to manage the customer's requests sent via the
\textit{UserInteraction} service. The \texttt{customerMenu}
\textit{pick} activity, contained in the body of the loop, may receive
four kinds of possible requests: three of them (\texttt{getBalance},
\texttt{deposit}, \texttt{withdraw}) are forwarded to the
corresponding operations of the \textit{BankAccount} service; the
\texttt{logOff} request terminates the loop, closing the customer
session.

\begin{figure}[htb]
\centering
\includegraphics[width=\linewidth]{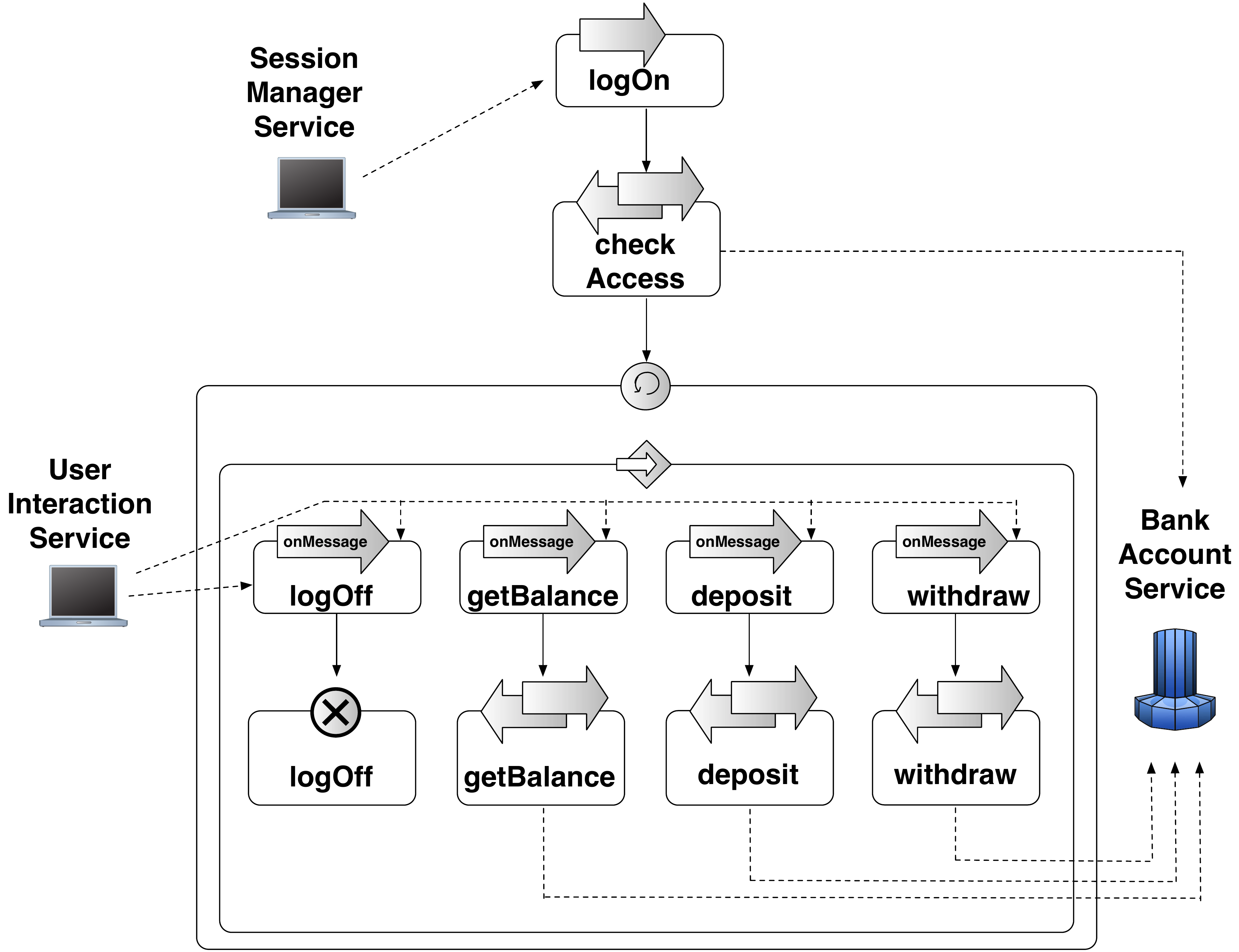}
\caption{\emph{ATMFrontEnd} business process}
\label{fig:process}
\end{figure}

To annotate a BPEL process with \sol, 
 we denote the execution of each activity with an atom. 
For example, \textit{receive} activity \texttt{logOn} can be denoted with
an atom $\mathit{logOn} \in \Pi$.
 Synchronous \emph{invoke}
activities are actually modeled with two atoms, corresponding to
the start and the end of the invocation; these are denoted with the ``\_start'' and 
``\_end'' suffixes, respectively.

Below we list some examples of quantitative properties expressed first
in natural language and then with \sol; more details on the features of the
language are available in~\cite{bianculli13:_tale_solois}.  All
properties are under the scope of an implicit universal temporal
quantification as in \emph{``In every process run, \ldots''}; we
assume the time units to be in seconds.
\begin{enumerate}[QP1:]
\item \textsf{WithdrawalLimit}\\
  The number of withdrawal operations
  performed within 10 minutes before customer logs off is less than or equal to the allowed limit (assumed to be 3, for
example).
This property is expressed as:\\
$\mathsf{G}(\mathit{logOff} \rightarrow \mathfrak{C}^{600}_{\leq 3}(\mathit{withdraw}))$.
\item \textsf{CheckAccessAverageResponseTime} \\	
The average response time of operation \texttt{checkAccess} provided 
by the \textit{BankAccount}
service is always less than 5 seconds within any 15 minute time window. This property is expressed as:\\
$ \mathsf{G}(\mathfrak{D}^{900}_{< 5}\{(\mathit{checkAccess\_start},\mathit{checkAccess\_end})\})$.

\item \textsf{MaxNumberOfBalanceInquiries} \\
The maximum number of
balance inquiries is restricted to at most 2 
per minute within 10 minutes before customer session ends. This property is expressed as: \\ 
$\mathsf{G}(\mathit{logOff} \rightarrow \mathfrak{M}^{600,60}_{\leq 2}(\mathit{getBalance}))$.

\end{enumerate}

%%% Local Variables: 
%%% mode: latex
%%% TeX-master: "soca2014"
%%% End: 

\fi

\subsection{\ct}

\ct~\cite{bersani12:_const_ltl_satis_check_autom} is
an extension of PLTLB (Propositional Linear Temporal Logic with both
past and future modalities)~\cite{glory-pnulli} augmented with atomic
formulae built over a constraint system $\mathcal{D}$. In practice,
\ct defines a set of \emph{
% counter %PL:I'd say we call them variables and later we rename them as
%counters.  
variables} $C$ and
\emph{arithmetical constraints} over a constraint system
$\mathcal{D}$; 
in our case, $\mathcal{D}$ is the structure
$(\mathbb{Z},=,(<_{d})_{d \in \mathbb{Z}})$. For this particular combination, decidability of \ct has been proven
in~\cite{demri07:_ltl}. Each $<_d$ is a binary relation defined as $x
<_d y \Leftrightarrow x < y +d $, hence, for example, the notation $x= y+d$ is an
abbreviation for $y <_{1-d} x \land x <_{d+1} y$.
  Variables (henceforth called \emph{counters}) 
% are variables that 
  receive a separate evaluation at each time instant. In addition to
  the standard PLTLB temporal operators ``\emph{Since}'' and
  ``\emph{Until}'', \ct introduces the new construct
  of \emph{arithmetic temporal term}, defined as $\alpha := c \mid x
  \mid \mathsf{Y}(x) \mid \mathsf{X}(x)$, where $c \in \mathbb{Z}$ is
  a constant, $x \in C$ is a counter and $\mathsf{Y}$ and $\mathsf{X}$
  are temporal operators applied to counters. These temporal operators
  for counters return the value of
  the counter in the previous and in the next time instant,
  respectively.  Note that we use a syntactically sugared version of
  PLTLB using metric temporal operators over time intervals, such as $
  \mathsf{U}_I$.  Since time is discrete, they are just a convenient
  shorthand~\cite{pradella2007:the-symmetry-of}.  The
  syntax of \ct is the following:\\
$\phi \Coloneqq p \mid \alpha \sim \alpha \mid 
 \neg \phi \mid \phi \land \phi \mid \phi \mathsf{U}_I \phi \mid \phi
\mathsf{S}_I \phi \mid \mathsf{X} \phi \mid \mathsf{Y} \phi$\\
where $p$ is an atomic proposition, $\sim \in \{=,(<_{d})_{d \in
  \mathbb{Z}} \}$, $\mathsf{S}_I$, $\mathsf{U}_I$, $\mathsf{X}$, $\mathsf{Y}$ are
the usual ``\emph{Since}'', ``\emph{Until}'', ``\emph{Next}'', and
``\emph{Yesterday}'' modalities of  PLTLB. Additional temporal
modalities (like $\mathsf{G}$, ``Globally'', and $\mathsf{W}$, ``Weak Until'') can be
defined using the usual conventions. An example of a
\ct formula is $\mathsf{G}(\phi \rightarrow
\mathsf{X}(y)=y+1)$, which states that whenever $\phi$ is true, the
value of counter $y$ in the next time instant must be 
incremented of 1 with respect to the value at the current time instant.

\ct formulae admit finite, ultimately periodic
two-part models ($\pi$,$\delta$).  Function $\pi:\mathbb{N}
\rightarrow \mathcal{P}(\Pi)$ associates a subset of the 
propositions with each time instant, while function
$\delta:\mathbb{N}\times C \rightarrow \mathbb{Z}$ defines the value
of counters at each time position.  Hereafter, this two-part model
will be graphically represented as in Fig.~\ref{fig:dmod}: the
topmost row (above the timeline) represents function $\pi$ (e.g., $\pi(5)=\{\psi\}$); the
rows of integers below the timeline represent function $\delta$,
i.e., the values of each counter defined in the model. In the example
in the figure there are six counters, as shown on the left: $c_{\chi}, g_{\phi,\psi},
h_{\phi,\psi}, s_{\phi,\psi}, a_{\phi,\psi}, b_{\phi,\psi}$; the $\delta$
function is defined so that we have, for example in correspondence
with the sixth time instant (position \#5),  $\delta(5, g_{\phi,\psi})
= 1$, $\delta(5, h_{\phi,\psi}) = 0$, $\delta(5, s_{\phi,\psi}) = 3$, $\delta(5,
a_{\phi,\psi}) = 0$, and $\delta(5, b_{\phi,\psi}) = 3$.

%%% Local Variables: 
%%% mode: latex
%%% TeX-master: "soca2014"
%%% End: 

\section{The Translation from \sol to \ct}
\label{sec:basic-translation}
 
\begin{figure}[tb]
\footnotesize
 \begin{center}
 \ifreport
\begin{tikzpicture}[scale=3.2]
\else
\begin{tikzpicture}[scale=3]
\fi

%length of the line
\ifreport
\def \length{4}
\else
\def \length{2.8}
\fi
%y-position of the line
\def \heightI{0}
\def \heightII{-0.6}
%size of the circles
\def \circlesize{0.7pt}
%number of circles
\def \circlenumber{15}
%arrays with content
\def \eventsI{{"","",1,"","","","",2,"","",3,"","",4,""}}
\def \eventsII{{"","",3,"","","","",5,"","",3,"","",3,""}}
\def \eventsIII{{"","","$\chi$","","","","","$\phi$","","","$\phi$","","","$\psi$",""}}
\def \eventsIV{{"","","","","","","","$\psi$","","","$\chi$","","","$\chi$",""}}
\def \eventsV{{"$\neg e$","$\neg e$","$e$","$\neg e$","$\neg e$","$\neg e$","$\neg e$","$e$","$\neg e$","$\neg e$","$e$","$\neg e$","$\neg e$","$e$","$\neg e$"}}

\def \eventsVI{{0,0,1,0,0,0,0,1,0,0,1,0,0,1,0}}

\tkzDefPoint(0,\heightI){O} \tkzDefPoint(\length,\heightI){A}
\tkzDefPoint(0,\heightII){B} \tkzDefPoint(\length,\heightII){C}

\begin{scope}[very thick,decoration={
    markings,
    mark=at position 1 with {\arrow[scale=2]{>}}}
    ] 
    \tkzDrawSegments[postaction={decorate}](O,A) 

\end{scope}

\begin{scope}[very thick,decoration={
    markings,
    mark=at position 1 with {\arrow[scale=2]{>}}}
    ] 
    \tkzDrawSegments[postaction={decorate}](B,C) 

\end{scope}

\pgfmathsetmacro{\endl}{\circlenumber-1}
\pgfmathsetmacro{\seglength}{\length/(\circlenumber+1)}

 \foreach \i in {0,...,\endl}{
 
\pgfmathsetmacro{\xcoord}{(\i+1)*\seglength}
\tkzDefPoint(\xcoord,\heightI){Ph}
\tkzDefPoint(\xcoord,\heightII){Pl}

%different y coord

\tkzDefPoint(\xcoord, \heightII+0.1){PlT}

\tkzDefPoint(\xcoord,\heightI-0.1){Ph1}
\tkzDefPoint(\xcoord,\heightI+0.11){Ph2}
\tkzDefPoint(\xcoord,\heightI+0.22){Ph3}

\tkzDefPoint(\xcoord,\heightII-0.1){Pl1}
\tkzDefPoint(\xcoord,\heightII+0.21){Pl2}
\tkzDefPoint(\xcoord,\heightII+0.32){Pl3}

\tikzstyle{point}=[rectangle,fill=black,draw=black,minimum size=5pt,inner sep=0pt]

%circles
\pgfmathparse{\eventsVI[\i]}
\pgfsetfillopacity{\pgfmathresult}
\pgfsetstrokeopacity{\pgfmathresult}
\node [point] (0) at (Ph) {};
\pgfsetfillopacity{1}
\pgfsetstrokeopacity{1}

\filldraw (Pl) circle (\circlesize);

%numbering time instants
\node at (PlT) {\i};

%rows
\pgfmathparse{\eventsII[\i]}
\node at (Ph1) {\pgfmathresult};
\pgfmathparse{\eventsIII[\i]}
\node at (Ph2) {\pgfmathresult};
\pgfmathparse{\eventsIV[\i]}
\node at (Ph3) {\pgfmathresult};

\pgfmathparse{\eventsV[\i]}
\node at (Pl1) {\pgfmathresult};
\pgfmathparse{\eventsIII[\i]}
\node at (Pl2) {\pgfmathresult};
\pgfmathparse{\eventsIV[\i]}
\node at (Pl3) {\pgfmathresult};

}

\end{tikzpicture}

%%% Local Variables: 
%%% mode: latex
%%% TeX-master: "../soca2014"
%%% End: 
 \end{center}
 \caption{Mapping a timed $\omega$-word into an $\omega$-word}
 \label{fig:omega}
 \end{figure}
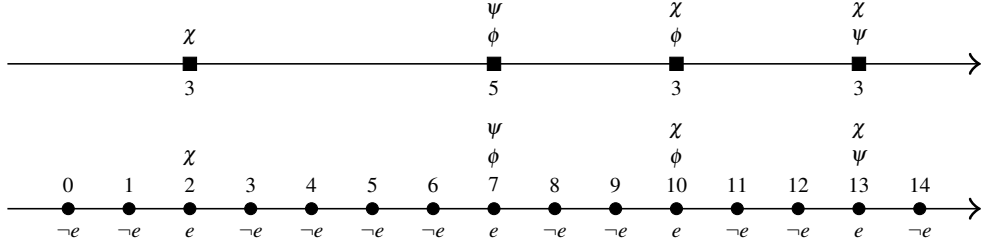

The key point in defining the translation from \sol to
\ct is to bridge the gap between the semantics of
\sol based on {\em timed} $\omega$-words, where the temporal information is
denoted by an integer time-stamp, and the one of \ct,
where the temporal information is implicitly defined by the integer
position in an $\omega$-word. The two temporal models can be
transformed into each other. 
%
%Here we are interested in pinpointing the
%positions in a \ct $\omega$-word that correspond to
%time-stamps where events occur in a \sol timed $\omega$-word.  
Here we are interested in pinpointing, in a \ct $\omega$-word, only the positions that correspond 
to actual time-stamps in a \sol timed $\omega$-word. These timestamps correspond
to instants where some event actually occurs.
To do
so, we add to the set $\Pi$ a special propositional symbol $e$, which
is true in each position corresponding to a ``valid'' time-stamp in
the timed $\omega$-word; a ``valid'' time-stamp is one where at least
an event, represented by a propositional symbol, occurs.  
An example
of this conversion is shown in Fig.~\ref{fig:omega}, where a timed
$\omega$-word is depicted in the timeline at the top and its
equivalent $\omega$-word corresponds to the timeline at the bottom;
notice the special symbols $\neg e$ that hold in positions in the
$\omega$-word which do not correspond to a ``valid'' time-stamp in the
timed $\omega$-word. 
% As shown in Fig.~\ref{fig:cmod}, we depict
% traces as monoinfinite timelines with filled circles representing time
% instants of $\omega$-words 
%and filled squares representing timed
%$\omega$-words.  
Hereafter, when displaying $\omega$-words, we will
omit the symbol $e$ from positions in the timeline, since its presence
can be implied by the presence of other propositional symbols in the
same position in the timeline.

 To define the translation from \sol to \ct we
 consider, without loss of expressiveness, only formulae in positive
 normal form, i.e., where negation may only occur on atoms (see, for
 example,~\cite{pradella2007:the-symmetry-of}).  First, we extend the
 syntax of the language by introducing a dual version for each
 operator in the original syntax, except for the
 $\mathfrak{C}^K_{\bowtie n}, \mathfrak{U}^{K,h}_{\bowtie n},
 \mathfrak{M}^{K,h}_{\bowtie n}, \mathfrak{D}^K_{\bowtie n}$
 modalities\footnote{A negation in front of one of the
   $\mathfrak{C}^K_{\bowtie n}, \mathfrak{U}^{K,h}_{\bowtie n},
   \mathfrak{M}^{K,h}_{\bowtie n}, \mathfrak{D}^K_{\bowtie n}$
   modalities becomes a negation of the relation denoted by the
   $\bowtie$ symbol, hence no dual version is needed for them.}: the
 dual of $\land$ is $\lor$; the dual of $\mathsf{U}_I$ is
 ``\emph{Release}'' $\mathsf{R}_I$: $\phi \mathsf{R}_I \psi \equiv
 \neg (\neg \phi \mathsf{U}_I \neg \psi)$; the dual of $\mathsf{S}_I$
 is ``\emph{Trigger}'' $\mathsf{T}_I$: $\phi \mathsf{T}_I \psi \equiv
 \neg (\neg \phi \mathsf{S}_I \neg \psi)$.  A formula is in
 \emph{positive normal form} if its alphabet is $\{\land, \lor,
 \mathsf{U}_I, \mathsf{R}_I, \mathsf{S}_I, \mathsf{T}_I,
 \mathfrak{C}^K_{\bowtie n}, \mathfrak{U}^{K,h}_{\bowtie n},
 \mathfrak{M}^{K,h}_{\bowtie n}, \mathfrak{D}^K_{\bowtie n} \} \cup
 \Pi \cup \bar{\Pi}$, where $\bar{\Pi}$ is the set of formulae of the
 form $\neg p$ for $p \in \Pi$.

We can now illustrate the translation $\rho$ from  \sol formulae to \ct. 
For the propositional ($\neg$, $\land$ and $\lor$) and temporal part ($\mathsf{U}_I$, $\mathsf{S}_I$, $\mathsf{R}_I$ and $\mathsf{T}_I$) of \sol the translation is straightforward:\vspace{0.5cm}\\ 
\ifreport
\vspace{-0.5cm}
\begin{align*}
\rho(p) &\equiv p, p \in \Pi
\\
\rho(\neg p) &\equiv \neg p, p \in \Pi
\\ 
\rho(\phi \land \psi) &\equiv \rho(\phi) \land \rho(\psi)
\\
\rho(\phi \lor \psi) &\equiv \rho(\phi) \lor \rho(\psi)
\\
\rho(\phi \mathsf{U}_I \psi) &\equiv (\neg e \lor \rho(\phi)) \mathsf{U}_I (e \land \rho(\psi))
\\
\rho(\phi \mathsf{S}_I \psi)  &\equiv (\neg e \lor \rho(\phi)) \mathsf{S}_I (e \land \rho(\psi))
\\
\rho(\phi \mathsf{R}_I \psi)  &\equiv (e \land \rho(\phi)) \mathsf{R}_I (\neg e \lor \rho(\psi))
\\
\rho(\phi \mathsf{T}_I \psi)  &\equiv (e \land \rho(\phi)) \mathsf{T}_I (\neg e \lor \rho(\psi))
\end{align*}
\else
%\begin{equation}
%\begin{array}{c c c}
%\centering
$\rho(p) \equiv p, p \in \Pi$;\hspace{1.28cm}$\rho(\phi \mathsf{U}_I \psi) \equiv (\neg e \lor \rho(\phi)) \mathsf{U}_I (e \land \rho(\psi))$\\
$\rho(\neg p) \equiv \neg p, p \in \Pi$;\hspace{0.7cm} $\rho(\phi \mathsf{S}_I \psi)  \equiv (\neg e \lor \rho(\phi)) \mathsf{S}_I (e \land \rho(\psi))$\\
$\rho(\phi \land \psi) \equiv \rho(\phi) \land \rho(\psi)$; $\rho(\phi \mathsf{R}_I \psi)  \equiv (e \land \rho(\phi)) \mathsf{R}_I (\neg e \lor \rho(\psi))$\\
$\rho(\phi \lor \psi) \equiv \rho(\phi) \lor \rho(\psi)$; $\rho(\phi \mathsf{T}_I \psi)  \equiv (e \land \rho(\phi)) \mathsf{T}_I (\neg e \lor \rho(\psi))$\\
%\end{array}
%\end{equation}
\fi

%In the rest of this section we focus on the translation
%of the $\mathfrak{C}^K_{\bowtie n}, 
%\mathfrak{M}^{K,h}_{\bowtie n},
%\mathfrak{D}^K_{\bowtie n}$ modalities and then we briefly discuss its
%complexity. We omit the translation of the
%$\mathfrak{U}$ modality since it can be expressed in
%terms of the $\mathfrak{C}$ modality by exploiting the equivalence $
%\mathfrak{U}_{\bowtie n}^{K,h} (\phi) \equiv \mathfrak{C}_{\bowtie n \cdot \lfloor \frac{K}{h} \rfloor}^{\lfloor
%  \frac{K}{h} \rfloor \cdot h} (\phi)$.

\ifreport
In the rest of this section we focus on the translation
of the $\mathfrak{C}^K_{\bowtie n}$, $\mathfrak{U}^{K,h}_{\bowtie n}$, $\mathfrak{M}^{K,h}_{\bowtie n}$ 
and $\mathfrak{D}^K_{\bowtie n}$ modalities.
\else
In the rest of this section we focus on the translation
of the $\mathfrak{C}^K_{\bowtie n}$ 
and $\mathfrak{D}^K_{\bowtie n}$
modalities.
We omit the translation of the
$\mathfrak{U}$ and $\mathfrak{M}$ modalities since they can be expressed in
terms of the $\mathfrak{C}$ modality. Specifically, for the $\mathfrak{U}$ modality we exploit the equivalence $
\mathfrak{U}_{\bowtie n}^{K,h} (\phi) \equiv \mathfrak{C}_{\bowtie n \cdot \lfloor \frac{K}{h} \rfloor}^{\lfloor
  \frac{K}{h} \rfloor \cdot h} (\phi)$. As for the $\mathfrak{M}$ modality, the equivalence depends on the $\bowtie$ operator; for example, formula  $\mathfrak{M}_{< n}^{K,h}
(\phi)$
is equivalent to 
$
\left ( \bigwedge_{m=0}^{\lfloor \frac{K}{h} \rfloor - 1} \mathsf{Y}^{m \cdot
  h} (\  \mathfrak{C}^{h}_{< n} \phi 
 \right ) \land \left (  \mathsf{Y}^{\lfloor \frac{K}{h} \rfloor \cdot h} (
 \mathfrak{C}^{(K \bmod h)}_{< n} \phi  )
  \right ) 
$.
\fi

\subsection{Translation of the $\mathfrak{C}$ modality}
\label{subsec:cmod}

The $\mathfrak{C}$ modality expresses a bound on the number of
occurrences of a certain event in a given time window; it comes
natural to use the counters available in \ct for the
translation.  Indeed, for each sub-formula of the form
$\mathfrak{C}_{\bowtie n}^K (\chi)$, we introduce a counter
$c_{\chi}$, constrained by a set of \ct axioms,
detailed below. Informally, these axioms define the value of $c_{\chi}$ 
such that at each time position it captures the number of occurrences of 
event $\chi$ seen in the past:
\begin{compactenum}[{A}1)]
  \item \label{ca1} $c_\chi=0$
  \item \label{ca2} $\mathsf{G}((e \land \chi) \rightarrow \mathsf{X}(c_\chi)=c_\chi+1)$
  \item \label{ca3}  $\mathsf{G}((\neg e \lor \neg \chi) \rightarrow \mathsf{X}(c_\chi)=c_\chi)$
\end{compactenum}

Axiom~A\ref{ca1} initializes the counter to zero. Axiom~A\ref{ca2} 
states that if there is an occurrence of a
valid event $\chi$, (denoted by $e \land \chi$) the value of the 
counter $c_\chi$ in the next time instant is increased by one with respect to
the value at the current time instant.
Axiom~A\ref{ca3} refers to the opposite
situation, when either there is no occurrence of the event $\chi$ or
the time instant is not valid (i.e., $e$ does not hold in that time
instant). In this case, the value of the counter in the next time
instant must have the same value as in the current time
instant.  Both axioms~A\ref{ca2} and A\ref{ca3} have to hold at every
time instant, so they are in the scope of a \textit{globally} temporal
operator.

We can calculate the exact number of occurrences by subtracting the
values of the counter at the appropriate time instants; we explain
this through the example in Fig.~\ref{fig:dmod}, which depicts a short
trace of length 21 and the values assumed by the counter $c_{\chi}$
(in the first row) at each time instant, as determined by the
axioms. In the example, to evaluate the formula $\mathfrak{C}_{>
  1}^{K}(\chi)$ with $K=11$ at time instant $t=16$, we subtract from
the value of the counter $c_\chi$ at time instant $t+1=17$ (since we
want to consider a possible occurrence of $\chi$ at time instant $t$)
the value of the counter at time instant 6 (i.e., $t-(K-1)=16-(11-1)$,
which is 11 time instants in the past with respect to time instant
$t+1$); these values are enclosed in the figure with diamond markers.
The value resulting from the subtraction $6-1=5$ is then compared to
the specified bound ($5 > 1$). In symbols, this  can be written as
$\mathsf{X}(c_{\chi})-\mathsf{Y}^{10}(c_{\chi})>1$ evaluated at time
instant $t$.  This intuition is captured by the following
\ct formula, which generalizes the translation of a
\sol sub-formula of the form $\mathfrak{C}_{\bowtie n}^{K} (\chi)$ :
\begin{equation*}% \label{eq:cmod}
\rho\left(\mathfrak{C}_{\bowtie n}^{K} (\chi)\right) \equiv \mathsf{X}(c_{\chi}) - \mathsf{Y}^{K-1}(c_{\chi}) \bowtie n
\end{equation*}
Notice that the axioms are conjuncted with the resulting translation
of the \sol formula,
thus effectively constraining the behavior of all the counters of type $c_{\chi}$.

%\begin{figure}[tb]
%\centering
%\input{figures/cmod}
%\caption{Sample trace showing the counters used for the translation of the $\mathfrak{C}$ modality}
%\label{fig:cmod}
%\end{figure}

%%% Local Variables: 
%%% mode: latex
%%% TeX-master: "soca2014"
%%% End: 

\ifreport
\subsection{Translation of the $\mathfrak{U}$ modality}
\label{subsec:vmod}
The translation of the $\mathfrak{U}$ modality is
defined in terms of the $\mathfrak{C}$ modality; it  can then be defined as follows:
\begin{equation*}
\rho \left ( \mathfrak{U}_{\bowtie n}^{K,h} (\phi) \right ) \equiv
\rho \left (\mathfrak{C}_{\bowtie n \cdot \lfloor \frac{K}{h} \rfloor}^{\lfloor
  \frac{K}{h} \rfloor \cdot h} (\phi)\right )
\end{equation*}

This translation ignores the 
tail subinterval of the $\mathfrak{U}$ modality, which is consistent
with the \sol semantics~\cite{bianculli13:_tale_solois}.

\begin{figure}[t]
\begin{center}
\includegraphics[scale=1.1]{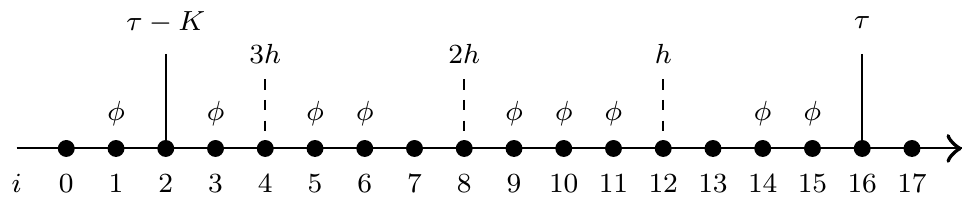}
\end{center}
\caption{Sample trace showing the time window and the observation
 subintervals considered for the evaluation of the
$\mathfrak{M}_{\geq 1}^{14,4}(\phi)$ formula at time instant $\tau=$16}
\label{fig:mmod}\end{figure}

%%% Local Variables: 
%%% mode: latex
%%% TeX-master: "soca2014"
%%% End: 

\subsection{Translation of the $\mathfrak{M}$ modality}
\label{subsec:mmod}
To translate the $\mathfrak{M}$ modality we rely on the
$\mathfrak{C}$ modality.
The translation of a formula of the form  $\mathfrak{M}_{< n}^{K,h}
(\phi)$ is defined as: $\rho\left(\mathfrak{M}_{< n}^{K,h} (\phi)\right) \equiv$
 \begin{equation*}
\left ( \bigwedge_{m=0}^{\lfloor \frac{K}{h} \rfloor - 1} \mathsf{Y}^{m \cdot
  h} (\ \rho ( \mathfrak{C}^{h}_{< n} \phi )
) \right ) \land \left (  \mathsf{Y}^{\lfloor \frac{K}{h} \rfloor \cdot h} (
 \ \rho ( \mathfrak{C}^{(K \bmod h)}_{< n} \phi  )
 ) \right ) 
 \end{equation*}
 
For a formula of the form $\mathfrak{M}_{> n}^{K,h} (\phi)$ we have:
$\rho\left(\mathfrak{M}_{> n}^{K,h} (\phi)\right) \equiv $ 
\begin{equation*}
\left(
\bigvee_{m=0}^{\lfloor \frac{K}{h} \rfloor - 1} \mathsf{Y}^{m \cdot h}
 (\ \rho ( \mathfrak{C}^{h}_{> n} \phi  )
 ) 
\right )
\lor
\left (
 \mathsf{Y}^{\lfloor \frac{K}{h} \rfloor \cdot h}  (
 \ \rho (  \mathfrak{C}^{(K \bmod h)}_{> n} \phi  )
 ) 
\right )
\end{equation*}
 
 The formula decomposes the computation of the maximum number of
 occurrences of the event $(e \land \phi)$ by suitably combining
 constraints on the number of occurrences of the event in each
 observation interval $h$ within the time window $K$.  The other cases of
 the operator $\bowtie$ can be defined in a similar way.

Fig.~\ref{fig:mmod} shows an example trace of length $18$. We evaluate the formula
$\mathfrak{M}_{\geq 3}^{14,4}(\phi)$ at time instant $\tau=16$. The vertical solid lines delimit the
time window of length $K=14$; the dashed lines delimit the adjacent non-overlapping observation
subintervals of length $h=4$.  The $\mathfrak{M}$ modality formula is translated into
a disjunction of four $\mathfrak{C}$ modality formulae each referring to a different subinterval. 
The first three ($\lfloor \frac{K}{h} \rfloor = \lfloor \frac{14}{4}
\rfloor = 3$) formulae have the form $\mathfrak{C}^{4}_{\geq 3}(\phi)$
and are evaluated at time instants $16(=16-0\cdot 4)$, $12(=16-1\cdot 4)$ and
$8(=16-2\cdot 4)$. The fourth formula (corresponding to rightmost disjunct defined in
the translation $\rho$) has the form $\mathfrak{C}^{2}_{\geq 3}(\phi)$ %(since $14 \bmod 4 = 2$), 
and is evaluated at time instant $4(=16 - \lfloor \frac{14}{4}\rfloor \cdot 4)$. We can conclude 
that, the formula $\mathfrak{M}_{\geq 3}^{14,4}(\phi)$ holds at time instant $\tau=16$ since formula 
$\mathfrak{C}^{4}_{\geq 3}(\phi)$ holds at time instant $12$ and renders the disjunction true. 

%%% Local Variables: 
%%% mode: latex
%%% TeX-master: "soca2014"
%%% End: 

\else
\fi

\subsection{Translation of the $\mathfrak{D}$ modality}
\label{subsec:dmod}
The $\mathfrak{D}$ modality expresses a bound on the average distance
between the occurrences of pairs of events in a given time
window. As anticipated in Sect.~\ref{sec:soloist-nutshell}, we consider only (sub)formulae of the $\mathfrak{D}$
modality that refer to one pair, like $\mathfrak{D}_{\bowtie n}^{K}(\phi, \psi)$.
%Before explaining the translation to \ct, we
%introduce some useful definitions. 

Events, corresponding to atomic propositions in \sol, can occur
multiple times in a trace; when we refer to a specific occurrence of
an event $\phi$ at a time instant $\tau$, we denote this as
$\phi_{|\tau}$.  Clearly, a pair of events $(\phi,\psi)$ may also have multiple
instances in a trace. We call a pair of the form $(\phi_{|i},\psi_{|j})$ an
\emph{instance} if there is an occurrence of event $\phi$ at time
instant $i$ and an occurrence of event $\psi$ at time instant $j$,
with $i<j$. We call such instance \textit{open} at time instant $\tau$
if $i\leq\tau<j$.  Otherwise, the instance is \textit{closed} at time
instant $\tau$.  The \emph{distance} of a closed (pair) instance is
$j-i$; for an open pair at time instant $\tau$, the distance is
$\tau-i$.  A time window of length $K$ defined for a $\mathfrak{D}$ modality
(sub-)formula evaluated at time instant $\tau$ is bounded by the time
instants $\tau+1$ and $\tau-K+1$.  For a certain trace, we say that a
$\mathfrak{D}$ modality (sub-)formula for a pair of events
$(\phi,\psi)$ has a \textit{left-open} pair in the trace if there is an open
instance of $(\phi,\psi)$ at time instant $\tau-K+1$ in the trace;
similarly, we say that the (sub-)formula has a \emph{right-open} pair
in the trace if there is an open instance of $(\phi,\psi)$ at time instant
$\tau+1$ in the trace.  The translation has then to take into account four
distinct cases, depending on whether a $\mathfrak{D}$ modality
(sub-)formula contains either (left- and/ or right-) open pairs or
none.

As done in the case of the  $\mathfrak{C}$ modality, the translation is based
on \ct counters.
For each sub-formula of the form  $\mathfrak{D}_{\bowtie n}^{K}(\phi,
\psi)$, we introduce
five counters, namely:
\begin{itemize}[-]
\item  $g_{\phi,\psi}$: this binary counter assumes value 1
  in the time instants following an occurrence of $\phi$ and it is reset
  to 0 after an  occurrence of $\psi$. It acts as a flag denoting the
  time instants during which the event pair instance is open;
\item $h_{\phi,\psi}$: in each time instant, this counter contains
  the number of previously-seen closed pair instances.  It is
  increased after every occurrence of $\psi$;
\item $s_{\phi,\psi}$: at each time instant, the
    value of this counter corresponds to  the sum of  distances of all previously
  occurred pair instances. It is increased at every time
  instant when either $g_{\phi,\psi}=1$ holds or  $\phi$ occurs;
\item $a_{\phi,\psi}$: this counter keeps track of the  sum of the  distances of all previously occurred closed pair instances;
\item $b_{\phi,\psi}$: this counter has the values
  that will be assumed by counter  $s_{\phi,\psi}$  at the next occurrence of
  $\psi$ (more details below).
\end{itemize}
Counters $a_{\phi,\psi}$, $b_{\phi,\psi}$, and $h_{\phi,\psi}$ are directly used
 in the translation of the $\mathfrak{D}$ modality (sub-)formulae, while
counters $g_{\phi,\psi}$ and $s_{\phi,\psi}$ are helper counters, used
to determine  the values of the other counters.
These five counters are  constrained by the following  axioms:
\begin{compactenum}[{A}1)]
\setcounter{enumi}{3}
  \item \label{da4}  $g_{\phi,\psi}=0$ $\land$ $h_{\phi,\psi}=0$ $\land$ $a_{\phi,\psi}=0$ $\land$ $s_{\phi,\psi}=0$
  \item \label{da5} $(\mathsf{X}(b_{\phi,\psi})=b_{\phi,\psi})\mathsf{W}(e \land \psi)$
  \item \label{da6} 
 $\mathsf{G}$ $((e \land \phi \land \neg\psi) \rightarrow (\mathsf{X}(g_{\phi,\psi})=1 \ \land 
  \mathsf{X}(s_{\phi,\psi})=s_{\phi,\psi}+1 \ \land $
$ \mathsf{X}(h_{\phi,\psi})=h_{\phi,\psi} \ \land 
  \mathsf{X}(a_{\phi,\psi})=a_{\phi,\psi}))$ 
  \item \label{da7}
 $\mathsf{G}$ $((e \land \psi \land \neg\phi) \rightarrow (\mathsf{X}(g_{\phi,\psi})=0 \ \land 
 \mathsf{X}(h_{\phi,\psi})=h_{\phi,\psi}+1 \ \land $ 
 $\mathsf{X}(a_{\phi,\psi})=s_{\phi,\psi} \ \land
 \mathsf{X}(s_{\phi,\psi})=s_{\phi,\psi} \land b_{\phi,\psi}=s_{\phi,\psi} \ \land $ 
 $\mathsf{X}((\mathsf{X}(b_{\phi,\psi})=b_{\phi,\psi})\mathsf{W}(e \land \psi))))$  
  \item \label{da8}
 $\mathsf{G}$ $((\neg e \lor (\neg \phi \land \neg \psi))
 \rightarrow (\mathsf{X}(g_{\phi,\psi})=g_{\phi,\psi} \ \land $ 
 $\mathsf{X}(h_{\phi,\psi})=h_{\phi,\psi}
  \land$ $\mathsf{X}(a_{\phi,\psi})=a_{\phi,\psi} \ \land$ 
 $(g_{\phi,\psi}=1$ $\rightarrow$ $\mathsf{X}(s_{\phi,\psi})=s_{\phi,\psi}+1 ) \ \land $
 $ (g_{\phi,\psi}=0$ $\rightarrow$ $\mathsf{X}(s_{\phi,\psi})=s_{\phi,\psi})))$ 
  \item \label{da9}
 $\mathsf{G}$ $((e \land \phi \land \psi)
 \rightarrow (\mathsf{X}(g_{\phi,\psi})=g_{\phi,\psi} \land
 \mathsf{X}(h_{\phi,\psi})=h_{\phi,\psi}+1 \ \land$
$\mathsf{X}(a_{\phi,\psi})=a_{\phi,\psi} \land \mathsf{X}(s_{\phi,\psi})=s_{\phi,\psi} \ \land$ 
 $\mathsf{X}((\mathsf{X}(b_{\phi,\psi})=b_{\phi,\psi})\mathsf{W}(e \land \psi)) $ 
  \end{compactenum}

  Axiom~A\ref{da4} initializes all counters except counter
  $b_{\phi,\psi}$, which will assume values determined by counter
  $s_{\phi,\psi}$.  Axiom~A\ref{da5} states that the value of counter
  $b_{\phi,\psi}$ will stay the same in all the time
  instants until the first occurrence of $\psi$. Notice that we use
  the $\mathsf{W}$ modality (``weak until''), to deal with traces
  without occurrences of $\psi$.  Axiom~A\ref{da6} determines the next
  time instant value of the following counters, upon occurrence of a
  $\phi$ and absence of a $\psi$ event (denoted by $e \land \phi \land \neg\psi$): 
  counter $g_{\phi,\psi}$ is
  set to 1; counter $s_{\phi,\psi}$ is incremented by 1; counters
  $h_{\phi,\psi}$ and $a_{\phi,\psi}$ are constrained not to change
  in the next time instant. Axiom~A\ref{da7} determines
  how the counters are updated when a $\psi$ event occurs and a $\phi$
  event does not: counter
  $g_{\phi,\psi}$ is set to 0; counters $b_{\phi,\psi}$,
  $\mathsf{X}a_{\phi,\psi}$, and $\mathsf{X}s_{\phi,\psi}$ are set to
  be equal to $s_{\phi,\psi}$. Moreover, a formula equivalent to axiom~A\ref{da5}
  holds in the next time instant, forcing the value of
  $b_{\phi,\psi}$ to stay the same in all the following time
  instants until the next occurrence of $\psi$.  Axiom~A\ref{da8}
  covers the cases either when there are no valid events or when
  neither $\phi$ nor $\psi$ occur.  In these cases the values of
  counters $g_{\phi,\psi}$, $h_{\phi,\psi}$, and $a_{\phi,\psi}$ are
  constrained to stay the same, while counter $b_{\phi,\psi}$ is
  unconstrained. As for counter $s_{\phi,\psi}$, we need to distinguish two
  separate cases:  when the pair instance is open (denoted by
  $g_{\phi,\psi}=1$), counter $s_{\phi,\psi}$ is incremented by 1,
  otherwise it stays the same. Axiom~A\ref{da9} handles the case when both 
  events $\phi$ and $\psi$ hold, by incrementing counter $h_{\phi,\psi}$ by 1 
  and constraining the value of counter $b_{\phi,\psi}$ in the same way like axiom~A\ref{da7}. 
  The values of the other counters are constrained to stay the same. 
 
\begin{figure}[t]
\centering
\footnotesize
\begin{tikzpicture}[scale=2.6]

%length of the line
\def \length{3.4}
%y-position of the line
\def \height{0}
%size of the circles
\def \circlesize{0.7pt}
%number of circles
\def \circlenumber{21}
%arrays with content
\def \events{{"","","$\phi$","","$\chi$","$\psi$","$\chi$","$\chi$","","$\phi$","$\chi$","","","$\chi$","$\psi$","$\chi$","","$\phi$","","$\psi$",""}}
\def \rowI{{0,0,0,1,1,1,0,0,0,0,1,1,1,1,1,0,0,0,1,1,0}}
\def \rowII{{0,0,0,0,0,0,1,1,1,1,1,1,1,1,1,2,2,2,2,2,3}}
\def \rowIII{{0,0,0,1,2,3,3,3,3,3,4,5,6,7,8,8,8,8,9,10,10}}
\def \rowIV{{0,0,0,0,0,0,3,3,3,3,3,3,3,3,3,8,8,8,8,8,10}}
\def \rowV{{3,3,3,3,3,3,8,8,8,8,8,8,8,8,8,10,10,10,10,10,""}}
\def \rowVI{{0,0,0,0,0,1,1,2,3,3,3,4,4,4,5,5,6,6,6,6,6}}

\tkzDefPoint(0,\height){O} \tkzDefPoint(\length,\height){A}

\begin{scope}[very thick,decoration={
    markings,
    mark=at position 1 with {\arrow[scale=2]{>}}}
    ] 
    \tkzDrawSegments[postaction={decorate}](O,A) 

\end{scope}

\pgfmathsetmacro{\endl}{\circlenumber-1}
\pgfmathsetmacro{\seglength}{\length/(\circlenumber+1)}

%counter names
\tkzDefPoint(0, -0.12){C1}
\node at (C1) {$c_{\chi}$};
\tkzDefPoint(0, -0.23){C1}
\node at (C1) {$g_{\phi,\psi}$};
\tkzDefPoint(0, -0.33){C1}
\node at (C1) {$h_{\phi,\psi}$};
\tkzDefPoint(0, -0.43){C1}
\node at (C1) {$s_{\phi,\psi}$};
\tkzDefPoint(0, -0.53){C1}
\node at (C1) {$a_{\phi,\psi}$};
\tkzDefPoint(0, -0.63){C1}
\node at (C1) {$b_{\phi,\psi}$};

 \foreach \i in {0,...,\endl}{ 
\pgfmathsetmacro{\xcoord}{(\i+1)*\seglength}
\tkzDefPoint(\xcoord,0){P}

%different y coord
\tkzDefPoint(\xcoord, 0.21){PE}
\tkzDefPoint(\xcoord, 0.1){PT}

\tkzDefPoint(\xcoord,-0.1){Pcc}

\tkzDefPoint(\xcoord,-0.22){P1}
\tkzDefPoint(\xcoord,-0.32){P2}
\tkzDefPoint(\xcoord,-0.42){P3}
\tkzDefPoint(\xcoord,-0.52){P4}
\tkzDefPoint(\xcoord,-0.62){P5}

%circles
\filldraw (P) circle (\circlesize);

%events
\pgfmathparse{\events[\i]}
\node at (PE) {\pgfmathresult};

%numbering time instants
\node at (PT) {\i};

%rows
\pgfmathparse{\rowVI[\i]}
\node at (Pcc) {\pgfmathresult};

\pgfmathparse{\rowI[\i]}
\node at (P1) {\pgfmathresult};

\pgfmathparse{\rowII[\i]}
\node at (P2) {\pgfmathresult};

\pgfmathparse{\rowIII[\i]}
\node at (P3) {\pgfmathresult};

\pgfmathparse{\rowIV[\i]}
\node at (P4) {\pgfmathresult};

\pgfmathparse{\rowV[\i]}
\node at (P5) {\pgfmathresult};
}

%diamond
\pgfmathsetmacro{\xcoord}{7*\seglength}
\tkzDefPoint(\xcoord,-0.1){CR}
\node[diamond,draw,very thick,color=red,minimum size=4mm] at (CR) {};

\pgfmathsetmacro{\xcoord}{18*\seglength}
\tkzDefPoint(\xcoord,-0.1){CR}
\node[diamond,draw,very thick,color=red,minimum size=4mm] at (CR) {};

%circles
\pgfmathsetmacro{\xcoord}{3*\seglength}
\tkzDefPoint(\xcoord,-0.32){CR}
\node[circle,draw,very thick,color=red,minimum size=4mm] at (CR) {};

\pgfmathsetmacro{\xcoord}{5*\seglength}
\tkzDefPoint(\xcoord,-0.32){CR}
\node[rectangle,draw,very thick,color=red,minimum size=4mm] at (CR) {};

\pgfmathsetmacro{\xcoord}{6*\seglength}
\tkzDefPoint(\xcoord,-0.33){CR}
\node[regular polygon, regular polygon sides=3, minimum size=4mm, draw,very thick,color=red] at (CR) {};

\pgfmathsetmacro{\xcoord}{8*\seglength}
\tkzDefPoint(\xcoord,-0.32){CR}
\node[regular polygon, regular polygon sides=6, minimum size=4mm, draw,very thick,color=red] at (CR) {};

\pgfmathsetmacro{\xcoord}{17*\seglength}
\tkzDefPoint(\xcoord,-0.32){CR}
\node[circle,draw,very thick,color=red,minimum size=4mm] at (CR) {};

\pgfmathsetmacro{\xcoord}{17*\seglength}
\tkzDefPoint(\xcoord,-0.32){CR}
\node[rectangle,draw,very thick,color=red,minimum size=4mm] at (CR) {};

\pgfmathsetmacro{\xcoord}{20*\seglength}
\tkzDefPoint(\xcoord,-0.33){CR}
\node[regular polygon, regular polygon sides=3, minimum size=4mm, draw,very thick,color=red] at (CR) {};

\pgfmathsetmacro{\xcoord}{20*\seglength}
\tkzDefPoint(\xcoord,-0.33){CR}
\node[regular polygon, regular polygon sides=6, minimum size=4mm, draw,very thick,color=red] at (CR) {};

\pgfmathsetmacro{\xcoord}{3*\seglength}
\tkzDefPoint(\xcoord,-0.52){CR}
\node[circle,draw,very thick,color=red,minimum size=4mm] at (CR) {};

\pgfmathsetmacro{\xcoord}{8*\seglength}
\tkzDefPoint(\xcoord,-0.52){CR}
\node[regular polygon, regular polygon sides=6, minimum size=4mm, draw,very thick,color=red] at (CR) {};

\pgfmathsetmacro{\xcoord}{17*\seglength}
\tkzDefPoint(\xcoord,-0.52){CR}
\node[circle,draw,very thick,color=red,minimum size=4mm] at (CR) {};

\pgfmathsetmacro{\xcoord}{17*\seglength}
\tkzDefPoint(\xcoord,-0.52){CR}
\node[rectangle,draw,very thick,color=red,minimum size=4mm] at (CR) {};

\pgfmathsetmacro{\xcoord}{20*\seglength}
\tkzDefPoint(\xcoord,-0.53){CR}
\node[regular polygon, regular polygon sides=3, minimum size=4mm, draw,very thick,color=red] at (CR) {};

\pgfmathsetmacro{\xcoord}{20*\seglength}
\tkzDefPoint(\xcoord,-0.53){CR}
\node[regular polygon, regular polygon sides=6, minimum size=4mm, draw,very thick,color=red] at (CR) {};

\pgfmathsetmacro{\xcoord}{5*\seglength}
\tkzDefPoint(\xcoord,-0.62){CR}
\node[rectangle,draw,very thick,color=red,minimum size=4mm] at (CR) {};

\pgfmathsetmacro{\xcoord}{6*\seglength}
\tkzDefPoint(\xcoord,-0.63){CR}
\node[regular polygon, regular polygon sides=3, minimum size=4mm, draw,very thick,color=red] at (CR) {};

\end{tikzpicture}
\caption{Sample trace showing the counters used for the translation of
  the $\mathfrak{C}$ and  $\mathfrak{D}$ modalities}
\label{fig:dmod}
\end{figure}
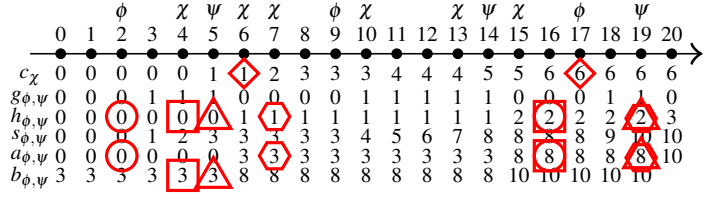

  As said above, the $b_{\phi,\psi}$ counter keeps the values that
  will be assumed by counter $s_{\phi,\psi}$ at the next occurrence of
  $\psi$.  The value assumed by both counters $a_{\phi,\psi}$ and
  $b_{\phi,\psi}$ originates from counter $s_{\phi,\psi}$, as enforced
  by axiom~A\ref{da7}. Axioms~A\ref{da6} and A\ref{da8} make sure the
  value of $s_{\phi,\psi}$ is propagated in the future via counter
  $a_{\phi,\psi}$, while axiom A\ref{da7} enables the
  propagation of this value in the past via counter $b_{\phi,\psi}$.
  We elaborate this through an example: Fig.~\ref{fig:dmod}
  represents a short trace with event $\psi$
  occurring at time instants 5, 14, and 19. Axiom~A\ref{da5} enforces equality 
  %(denoted with \tikz[baseline=-0.5ex]{ \draw[snake=zigzag,segment length=4pt] (0,0)-- (0.5,0); })
   between successive values of counter
  $b_{\phi,\psi}$ at adjacent time instants until the first occurrence
  of $\psi$ (time instants 0--5).  Additional equalities (of the same type) on the values
  of counter $b_{\phi,\psi}$ are enforced by axiom~A\ref{da7} (time instants 6--14 and 15--19). The
  same axiom also determines equality 
  %(denoted with \tikz[baseline=-0.5ex]{ \draw[snake=coil,segment length=4pt]  (0,-0.25) -- (0,0.25); })
   between the values of the
  $s_{\phi,\psi}$ and $b_{\phi,\psi}$ counters upon an occurrence of
  $\psi$ (time instants 5, 14 and 19).

The translation  $\rho\left(\mathfrak{D}_{\bowtie
  n}^{K}(\phi, \psi) \right)$ is defined as:
\begin{equation*}
\begin{array}{ll}
  \text{\texttt{if\footnotemark} }  (\mathsf{Y}^{K-1}  (g_{\phi,\psi}) = 1)
&
 \text{\texttt{then} }
(\tfrac{\mathsf{X}(a_{\phi,\psi})-\mathsf{Y}^{K-1}(b_{\phi,\psi})}{\mathsf{X}(h_{\phi,\psi})-\mathsf{Y}^{K-1}(h_{\phi,\psi})-1}
\bowtie n \land Z_1)
\\  & 
 \text{\texttt{else} }
(\tfrac{\mathsf{X}(a_{\phi,\psi})-\mathsf{Y}^{K-1}(a_{\phi,\psi})}{\mathsf{X}(h_{\phi,\psi})-\mathsf{Y}^{K-1}(h_{\phi,\psi})}
\bowtie n \land Z_2)
\end{array}
\end{equation*}
\footnotetext{``\texttt{if} $A$ \texttt{then} $B$ \texttt{else} $C$'' can be written as $(A \land B)\lor(\neg A \land C)$}

The condition $\mathsf{Y}^{K} (g_{\phi,\psi}) = 1$ checks whether the
time window contains an open pair instance on its left bound. Since
the semantics of the $\mathfrak{D}$ modality considers only closed
pairs within the time window to compute the average distance,  open
pairs must be ignored both on the left and on the right bound of the
time window.  There is no need to differentiate between the cases
when there is a right-open pair, since counter $a_{\phi,\psi}$
only considers  distances between closed pair instances.
The numerator of the fraction in both the \texttt{then} and
\texttt{else} branches denotes the total
distance, while the denominator corresponds to the number of pair
instances considered for computing the total distance.
Propositions $Z_1$ and $Z_2$ are respectively $\mathsf{X}(h_{\phi,\psi})-\mathsf{Y}^{K-1}(h_{\phi,\psi})\neq1$
and $\mathsf{X}(h_{\phi,\psi})-\mathsf{Y}^{K-1}(h_{\phi,\psi})\neq0$; due to these disjuncts 
the $\mathfrak{D}$ modality evaluates to true when there are no closed pairs
in the time window $K$.
Axioms~A\ref{da4}, A\ref{da5}, A\ref{da6}, A\ref{da7}, A\ref{da8}, A\ref{da9} are
conjuncted with the resulting translation and added as constraints
that hold at the initial time instant of the  trace.

An example of the use of counters to evaluate a formula with the $\mathfrak{D}$
modality is shown in Fig.~\ref{fig:dmod}, which depicts a simple 
trace and the values assumed by the counters  $g_{\phi,\psi}$,
$h_{\phi,\psi}$, $s_{\phi,\psi}$, $a_{\phi,\psi}$, and $b_{\phi,\psi}$
at each time instant, as
determined by the axioms. We notice that there are three 
instances of the  $(\phi, \psi)$ pair.
If we evaluate the formula
 $\mathfrak{D}_{\bowtie n}^{14}(\phi, \psi)$ at time instant 15, 
 the two pair instances
 $(\phi_{|2},\psi_{|5})$ and $(\phi_{|9},\psi_{|14})$,  considered
 to compute the average distance, are closed. The left-hand side (lhs)
 of
 the comparison operator ($\bowtie$) is evaluated  using the values of
counters $a_{\phi,\psi}$ and $h_{\phi,\psi}$ at time instants 16 and 2
(enclosed in a circle in the figure), resulting in $\frac{8}{2}=4$.
When the same formula is evaluated at time instant 18, the portion of
the  trace considered contains both a left-open
$(\phi_{|2},\psi_{|5})$ pair and a right-open
$(\phi_{|17},\psi_{|19})$ one. The lhs of the comparison operator is
evaluated using the values of counters  $a_{\phi,\psi}$,
$b_{\phi,\psi}$, and $h_{\phi,\psi}$  at time instants 19 and 5
(enclosed in  a triangle in the figure ); its value is $\frac{5}{1}=5$.
Now consider the formula  $\mathfrak{D}_{\bowtie n}^{12}(\phi,
\psi)$. When evaluated at time instant 15, it has a left-open pair
$(\phi_{|2},\psi_{|5})$. The values of the counters $a_{\phi,\psi}$,
$b_{\phi,\psi}$, and $h_{\phi,\psi}$ considered to compute the lhs of
the comparison operator are those at time instants 16 and 4 (enclosed
in a square in the figure); the lhs evaluates to $\frac{5}{1}=5$. If
the same formula is evaluated at time instant 18, we find only a
right-open pair $(\phi_{|17},\psi_{|19})$. The lhs of the comparison
operator is evaluated using the value of counters $a_{\phi,\psi}$ and
$h_{\phi,\psi}$ considered at time instants 19 and 7 (enclosed in a
hexagon in the figure); its value is $\frac{5}{1}=5$.

%%% Local Variables: 
%%% mode: latex
%%% TeX-master: "soca2014"
%%% End: 

%\input{table}

% \subsection{Complexity of the translation}
%  The translation function $\rho$, for the atomic
% propositions, the temporal modalities and all the aggregate one but
% $\mathfrak{M}$, introduces a fixed-length formula; notice that
% subformulae occuring in aggregate modalities are restricted to be
% atomic. The translation of the $\mathfrak{M}$ modality is linear with
% respect to the length $K$ of the time window.  In the worst case, our
% translation has a quadratic complexity, given that it is linear in the size 
% of the input formula and in the maximum constant $K$
% occurring in $\mathfrak{M}$ modalities.

\subsection{Complexity of the translation}
 The translation function $\rho$, for the atomic
propositions, the temporal modalities and all the
 aggregate ones, introduces a fixed-length formula; 
 notice that subformulae occurring in aggregate 
 modalities are restricted to be
atomic.  In the worst case, our translation is 
linear in the size of the input formula. We remark 
that we use a direct encoding of the exponent $K$ 
in formulae of the form $\mathsf{Y}^K$ or $\mathsf{X}^K$, 
both in the case of arithmetical temporal terms and 
of boolean formulae. The direct encoding of the exponent 
allows us to avoid expanding it into nested $\mathsf{Y}$ 
or $\mathsf{X}$ formulae.

%%% Local Variables: 
%%% mode: latex
%%% TeX-master: "soca2014"
%%% End: 

\section{Implementation}
\label{sec:implementation}

The translation described in the previous section has been implemented
in a tool~\cite{krstic13:_solois_trans}; 
this tool acts as a front-end
for translating \sol formulae into the input format of the ZOT
verification toolset~\cite{pradella2013:bounded-satisfi}.  ZOT
supports satisfiability checking of \ct formulae by
means of SMT solvers. A plugin-based architecture makes it easy to extend ZOT
to support more expressive languages using \ct as a
core, and to output code for the different dialects of various SMT
solvers.  We implemented the support for \sol as a ZOT plugin written
in Common Lisp.

%Refers to servics

%We use ZOT (together with the plugin we developed) to perform \emph{offline trace
%checking} of service execution traces against requirements expressed in
%\sol. As anticipated in Sect.~\ref{sec:introduction}, 
%in the context of this paper we assume that a trace is composed by the events
%corresponding to the interactions of a composite service with its
%external services. We express this execution trace also as a temporal
%logic formula; at each time instant, an atomic proposition holds if
%and only if the corresponding service interaction event (e.g., the invocation of an operation on a partner
%service) occurred.

\ifreport
\begin{figure}
\else
\begin{figure}[b]
\fi

\centering

\ifreport
\newcolumntype{A}{>{\centering\arraybackslash}m{5cm}}
\newcolumntype{B}{>{\raggedright\arraybackslash}m{1.2cm}}
\newcolumntype{C}{>{\centering\arraybackslash}m{2.2cm}}
\newcolumntype{G}{>{\centering\arraybackslash}m{2.8cm}}
\footnotesize	

\else
\newcolumntype{A}{>{\centering\arraybackslash}m{3.4cm}}
\newcolumntype{B}{>{\raggedright\arraybackslash}m{0.7cm}}
\newcolumntype{C}{>{\centering\arraybackslash}m{1.9cm}}
\newcolumntype{G}{>{\centering\arraybackslash}m{1.5cm}}

\tiny
\fi

\setlength{\tabcolsep}{5pt}
\begin{tabular}{ | B | C | G | A | }
\hline
&
 Trace&
Formula&
Counter constraints\\
\hline
\sol&
\ifreport
\hspace{0.1cm}{\scriptsize \begin{tikzpicture}[scale=1.5]

%length of the line
\def \length{1.3}
%y-position of the line
\def \heightI{0}
%size of the circles
\def \circlesize{0.7pt}
%number of circles
\def \circlenumber{7}
%arrays with content

\def \eventsII{{"","",2,"","","3","","","43","1"}}
\def \eventsIII{{"","","$p$","","","$p$","","","$p$","$p$"}}
\def \eventsVI{{0,0,1,0,0,1,0,0,1,1}}

\pgfmathsetmacro{\endl}{\circlenumber-1}
\pgfmathsetmacro{\seglength}{\length/(\circlenumber+1)}

\tkzDefPoint(0,\heightI){O} 
\tkzDefPoint(8*\seglength,\heightI){B} 
\tkzDefPoint(8*\seglength,\heightI){C} 
\tkzDefPoint(\length,\heightI){A}

%\begin{scope}[very thick
%    ] 
%    \tkzDrawSegments[](O,B) 
%
%\end{scope}

\begin{scope}[very thick,decoration={
    markings,
    mark=at position 1 with {\arrow[scale=1]{>}}}
    ] 
    \tkzDrawSegments[postaction={decorate}](O,B) 

\end{scope}

 \foreach \i in {0,...,\endl}{
 
\pgfmathsetmacro{\xcoord}{(\i+1)*\seglength}
\tkzDefPoint(\xcoord,\heightI){Ph}

%different y coord

\tkzDefPoint(\xcoord,\heightI-0.1){Ph1}
\tkzDefPoint(\xcoord,\heightI+0.11){Ph2}
\tkzDefPoint(\xcoord,\heightI+0.22){Ph3}

\tikzstyle{point}=[rectangle,fill=black,draw=black,minimum size=2pt,inner sep=0pt]

%circles
\pgfmathparse{\eventsVI[\i]}
\pgfsetfillopacity{\pgfmathresult}
\pgfsetstrokeopacity{\pgfmathresult}
\node [point] (0) at (Ph) {};
\pgfsetfillopacity{1}
\pgfsetstrokeopacity{1}

%rows
\pgfmathparse{\eventsII[\i]}
\node at (Ph1) {\pgfmathresult};
\pgfmathparse{\eventsIII[\i]}
\node at (Ph2) {\pgfmathresult};

}

\tkzDefPoint(7.55*\seglength,\heightI-0.004){T} 
%\node at (T) {$\cdots$};

\end{tikzpicture}

%%% Local Variables: 
%%% mode: latex
%%% TeX-master: "../icse2014"
%%% End: 
} &
\fi
$\mathfrak{C}^{5}_{< 3}(p)$
& n/a
\\
\hline
\ct&
\ifreport
\hspace{0.1cm}{\scriptsize \begin{tikzpicture}[scale=1.5]

%length of the line
\def \length{1.3}
%y-position of the line
\def \heightI{0}
\def \heightII{0}
%size of the circles
\def \circlesize{0.7pt}
%number of circles
\def \circlenumber{7}
%arrays with content
\def \eventsI{{0,1,2,3,4,5,6,0,48,49}}
\def \eventsIII{{"","","$p$","","","$p$","","","$p$","$p$"}}
\def \eventsIV{{"","","","","","","","","",""}}
\def \eventsV{{"$\neg e$","$\neg e$","$e$","$\neg e$","$\neg e$","$e$","$\neg e$","","$e$","$e$"}}
\def \eventsVI{{1,1,1,1,1,1,1,0,1,1}}

\pgfmathsetmacro{\endl}{\circlenumber-1}
\pgfmathsetmacro{\seglength}{\length/(\circlenumber+1)}

\tkzDefPoint(0,\heightII){B} 
\tkzDefPoint(8*\seglength,\heightII){A} 
\tkzDefPoint(8*\seglength,\heightII){O} 
\tkzDefPoint(\length,\heightII){C}

%\begin{scope}[very thick
%    ] 
%    \tkzDrawSegments[](B,A) 
%
%\end{scope}

\begin{scope}[very thick,decoration={
    markings,
    mark=at position 1 with {\arrow[scale=1]{>}}}
    ] 
    \tkzDrawSegments[postaction={decorate}](B,A) 

\end{scope}

 \foreach \i in {0,...,\endl}{
 
\pgfmathsetmacro{\xcoord}{(\i+1)*\seglength}

\tkzDefPoint(\xcoord,\heightII){Pl}
\tkzDefPoint(\xcoord, \heightII+0.1){PlT}
\tkzDefPoint(\xcoord,\heightII-0.1){Pl1}
\tkzDefPoint(\xcoord,\heightII+0.21){Pl2}

\pgfmathparse{\eventsVI[\i]}
\pgfsetfillopacity{\pgfmathresult}
\pgfsetstrokeopacity{\pgfmathresult}
\filldraw (Pl) circle (\circlesize);
\pgfmathparse{\eventsI[\i]}
\node at (PlT) {\pgfmathresult};
\pgfsetfillopacity{1}
\pgfsetstrokeopacity{1}

%numbering time instants

%rows

\pgfmathparse{\eventsV[\i]}
\node at (Pl1) {\pgfmathresult};
\pgfmathparse{\eventsIII[\i]}
\node at (Pl2) {\pgfmathresult};

}

\tkzDefPoint(7.55*\seglength,\heightII-0.004){T} 
%\node at (T) {$\cdots$};

\end{tikzpicture}

%%% Local Variables: 
%%% mode: latex
%%% TeX-master: "../icse2014"
%%% End: 
} &
\fi

$\underbrace{\underbrace{\underbrace{\mathsf{X}(c_p)}_{\text{a}}-\underbrace{\mathsf{Y}^{4}(c_p)}_{\text{b}}}_{\text{c}}<3}_{\text{d}}$
&
%\vspace{0.2cm}
$ Cc_p \left\{ \begin{array}{cc}
(c_p=0) &(A1)\\
\land&\\ 
\mathsf{G}((e \land p) \rightarrow \mathsf{X}(c_p)=c_p+1) & (A2)\\
\land&\\
\mathsf{G}((\neg e \lor \neg p) \rightarrow \mathsf{X}(c_p)=c_p) \hspace{0.01cm} & (A3)\\
\end{array} \right. $ 
%\vspace{0.2cm}
\\
\hline
SMT \newline input \newline language &
%\vspace{0.2cm}
(and \hspace{1cm}

(not ($e$ 0))

(not ($e$ 1))

($e$ 2)($p$ 2)

(not ($e$ 3))

(not ($e$ 4))

($e$ 5)($p$ 5)

(not ($e$ 6))

&

\vspace{0.2cm}

(and\hspace{1cm}

(= ($a$ $i$) ($c_p$ (+ $i$ 1))) 

[$i=0 \ldots 5$]

%\vspace{0.2cm}

(= (b i) ($c_p$ (- i 4))) 

[$i=4 \ldots 6$]

%\vspace{0.2cm}

(= (c i) (- (a i) (b i))) 

[$i=0 \ldots 6$]

%\vspace{0.2cm}

\hspace{-0.1cm}(iff (d i) (< (c i) 3)) )

[$i=0 \ldots 6$]

&

(and  \hspace{2.85cm}

(iff ($Cc_p$ i) \hspace{2.3cm}

\hspace{0.48cm}(and (A1 i) (A2 i) (A3 i))) )

[$i=0 \ldots 6$]\hspace{2.3cm}

$\vdots$

\\
\hline

\end{tabular}

\caption{Example of the translation from \sol to \ct
  and then to the input language of the SMT solver}
\label{fig:intuition}

\end{figure}

%%% Local Variables: 
%%% mode: latex
%%% TeX-master: "soca2014"
%%% End: 

We now give a rundown of the translation steps applied
to an example, to provide a glimpse of the implementation of our
SMT-based trace checking algorithm. 
 These steps and the example are also sketched in
Fig.~\ref{fig:intuition} where: the top row shows (a
fragment of) the
example input  trace and the \sol formula to verify on the trace;
the middle row shows how the input trace is
transformed from timed $\omega$-word to $\omega$-word, the translation of the input
formula and the definition of the counter constraints as described in
Sect.~\ref{sec:basic-translation}; the bottom row shows how the 
trace, the input formula, and the counter constraints are
translated into the input language of the SMT solver.

Let us consider the problem of performing trace checking of the
formula $\phi \equiv \mathfrak{C}^{5}_{< 3}(p)$ over the trace $H$ of
length 7 depicted in Fig.~\ref{fig:intuition}; the formula is
evaluated at time instant 5.  As described in Sect.~\ref{subsec:cmod},
our plugin translates the \sol formula
$\phi$ into \ct as $\rho(\phi) \equiv
\mathsf{X}(c_p)-\mathsf{Y}^{4}(c_p) < 3$, where $c_p$ is a
counter. The behavior of this counter is constrained by the
conjunction of axioms A\ref{ca1}, A\ref{ca2}, and A\ref{ca3}, defined
as $\mathcal{C}_{c_p} \equiv (c_p=0) \land \mathsf{G}((e \land p)
\rightarrow \mathsf{X}(c_p)=c_p+1) \land \mathsf{G}((\neg e \lor \neg
p) \rightarrow \mathsf{X}(c_p)=c_p)$.  The next step is to invoke ZOT
to translate the input formula and the counter constraints into the
input language of the SMT solver.  First, ZOT parses the formula and
assigns a special proposition to each sub-formula in the input
formula; similarly, it also assigns an arithmetic proposition to each arithmetical
temporal term in the input formula. For example, as shown in
Fig.~\ref{fig:intuition}, the arithmetic propositions $a$ and $b$
correspond, respectively, to the arithmetical temporal terms
$\mathsf{X}(c_p)$ and $\mathsf{Y}^{4}(c_p)$; $c$ is an arithmetical
proposition holding the value of the
$\mathsf{X}(c_p)-\mathsf{Y}^{4}(c_p)$ arithmetic temporal term;
proposition $d$ corresponds to the entire input formula.  The values
of these auxiliary propositions are defined in each time instant $i= 0
\ldots 6$, according to their semantics.  The trace $H$ is also
encoded in the input language of the SMT solver and provided to it as
an assumption. The SMT solver is then fed with the translation,
performed by ZOT, of the \ct formula
$\neg(\mathsf{X}^{5}(\rho(\phi))) \land \mathcal{C}_{c_p}$. Notice
that the formula $\phi$ is negated; hence, it is satisfied by trace
$H$ if the SMT solver returns \emph{unsat}. The exponent $5$ in the
term $\mathsf{X}^{5}(\rho(\phi))$ is determined by the evaluation of
the formula fixed at time instant 5.  The details of the
translation from \ct to the input language of the SMT
solver (as sketched in the bottom row of Fig.~\ref{fig:intuition})
have been omitted since they are out of the scope of this work; for
them, we refer the reader to~\cite{pradella2013:bounded-satisfi}.

%%% Local Variables: 
%%% mode: latex
%%% TeX-master: "soca2014"
%%% End: 

\section{Evaluation}
\label{sec:evaluation}

We evaluated the effectiveness of our approach by investigating the
following research questions:
\begin{itemize}
\item RQ1: \emph{How does the proposed approach scale with respect to the
    various parameters (e.g., the length of the trace, the length of
    the time window $K$) involved in \sol trace checking?} (Sect.~\ref{subsec:evaluation-histlength})
\item RQ2: \emph{How does the proposed trace checking procedure for
    \sol based on \ct
    compare with the procedure  based on \qf~\cite{bbgks-fase2014}?}
  (Sect.~\ref{sec:comparison-with-qf})
\ifreport
\item RQ3: \emph{Can the proposed trace checking procedure, based
    on  \ct,  handle traces more efficiently than the procedure  used in our first
    implementation~\cite{bianculli13:_tale_solois,Krstic:Thesis:2012},
    based on LTL?} (Sect.~\ref{subsec:evaluation-encab})
\item RQ4: \emph{Can the proposed trace checking procedure be applied in a 
    realistic setting?} (Sect.~\ref{subsec:evaluation-running})
\else
\fi
\end{itemize}

Since there is no consolidated benchmark for service-based applications
(for which \sol was tailored), we decided to evaluate our approach
using  synthesized traces.
\ifreport
These traces were obtained using the Process Log Generator (PLG) 
tool~\cite{burattin11:_plg} on a model of the running example
from Sect.~\ref{sec:running}. This model was defined by specifying the workflow
structure, the duration of each synchronous \textit{invoke} activity,
the branching probabilities, and the error rates. Other activities (e.g., \textit{receive}) were
given 0 as duration; branching was used to create loops and simulate
the behavior of the \textit{pick} activity. The PLG tool is able to
synthesize logs of process invocations from its input model.
\else
All traces were synthesized with the Process
Log Generator (PLG) tool~\cite{burattin11:_plg}, starting from a model
of a realistic service composition (the ``Order Booking'' business
process distributed with the Oracle SOA Suite), comprising 37
activities.  This model was defined by
specifying the workflow structure, the duration of each activity
invoking an external service\footnote{Other activities were given 0 as
duration.}, the branching probabilities (to simulate loops), and the error
rates.
\fi 
\ifreport
  For each run of the trace checker, we recorded the memory usage, the 
 translation time,  and the SMT verification time.
\else
 For each run of the trace checker, we recorded the memory usage and the SMT verification time.
\fi
 The evaluation was performed on a PC equipped with a 2.0GHz Intel
 Core i7-2630QM processor, running GNU/Linux Ubuntu 12.10 64bit, with
 2GB RAM allocated for the verification tool. 
 We used the \textit{Z3}~\cite{moura08:_z3} SMT solver v.~4.3.1.

\subsection{RQ1: Scalability of the approach}
\label{subsec:evaluation-histlength}
To investigate RQ1, we
  considered the following parameters:
 \begin{compactdesc}
 \item[Trace length.] It represents the length of the
   synthesized  trace and the bound given to the SMT solver.
   The length of each synthesized trace depends on the duration of the
  activities invoking an external service as well as on 
the branching
   probabilities of the loop(s) in the process.
 \item[Length of the time window.] It is used in the aggregate
   modalities; it corresponds to the $K$ parameter.
 \item[Bound of the comparison operator.] It is used in the aggregate
   modalities; it corresponds to the $n$ parameter.
 \end{compactdesc}
 
We present only the  results of the evaluation
done for the $\mathfrak{C}$ and $\mathfrak{D}$ modalities, since they are the
keystones  of the translation. 
We synthesized 20000 different traces, of variable
length between 100 and 2000.
We checked the following properties on them:
$\mathfrak{C}^{100}_{> 30}(p)$, and 
$\mathfrak{D}^{100}_{> 30}(p,q)$, with propositions $p$ and $q$
corresponding to the
start and end events of a service invocation 
of the process.
The results of executing trace
checking for each of these two properties on the synthesized traces
are shown in 
\ifreport
Fig.~\ref{plot:scalability-ch}
and~\ref{plot:scalability-dh}; in each row, 
the left plot shows
the time for the translation and the one taken by the SMT solver, while
the right plot shows the memory usage.
\else
the top plot in Fig.~\ref{plots}; the plot shows
the time taken by the trace checking procedure for checking 
 both the $\mathfrak{C}$ properties 
(shown in black) and the $\mathfrak{D}$ one (shown in gray).
\fi
Each point in the plot represents 
an average value of 10 trace check runs on traces of the same length.
The plots provide an intuition of the growth rate of the resources usage with
respect to the length of the input trace. The memory usage 
for the respective properties yields a very similar plot; we omitted it 
for space reasons. The memory dedicated for the
evaluation of properties with the $\mathfrak{C}$ modality  was exhausted at 2200
time instances, requiring 2.1GB of memory and 40 seconds to solve.
For the evaluation of the properties with the $\mathfrak{D}$ modality, the maximum
number of time instances manageable before exhausting the preset
memory limit was
2000. The lower value with respect to the $\mathfrak{C}$
modality is due to the linear multivariate constraints introduced in
the translation of the $\mathfrak{D}$ modality; these constraints are
harder to solve than the univariate one used for the
$\mathfrak{C}$ modality.

\ifreport

\setcounter{figure}{7}
\newcommand{\plots}{0.75}

\begin{figure*}
\pgfplotsset{yticklabel style={text width=2.4em,align=right}}
\pgfplotsset{xticklabel style={text height=1em,align=right}}
\centering

\begin{tabularx}{\textwidth-0.7cm}{X X X }

\vspace{-3cm}\subcaption{Comparison between  LTL and CLTLB($\mathcal{D}$)\label{plot:ltl}}&

\includegraphics[scale=\plots]{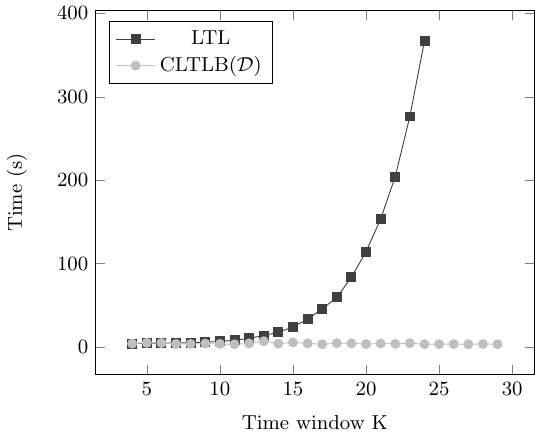}&

\includegraphics[scale=\plots]{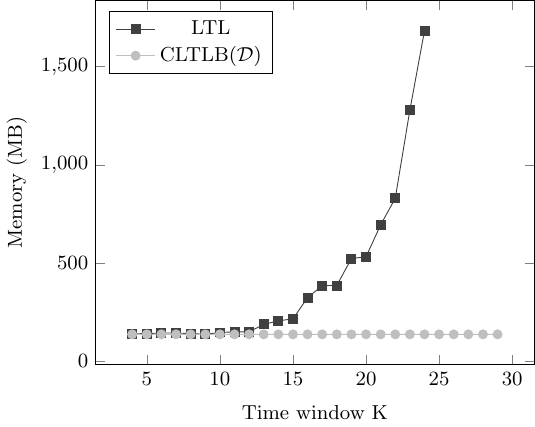}\\

\vspace{-3cm}\subcaption{Comparison between  QF-EUFIDL and CLTLB($\mathcal{D}$)
over traces with 100\%, 50\%, 33\%, 25\%, 20\%, 16.6\%, and 14.3\%  sparseness}\label{plot:qfeufidl}&

\includegraphics[scale=\plots]{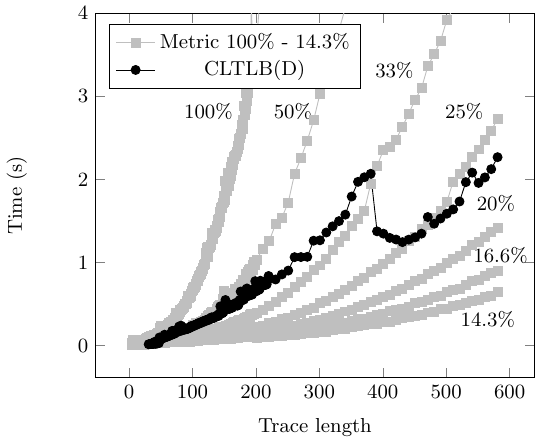}&

\includegraphics[scale=\plots]{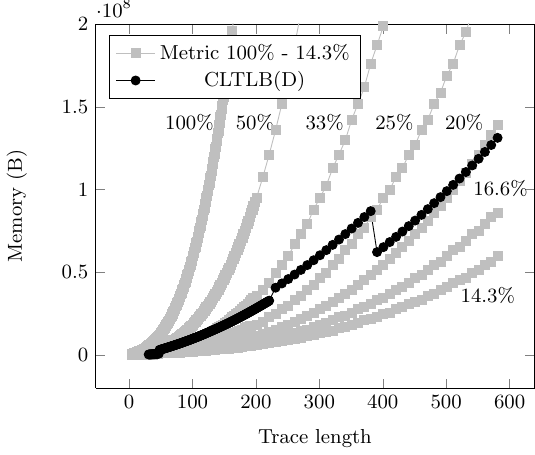}\\

\vspace{-3cm}\subcaption{Time and memory scalability of  $\mathfrak{C}$ modality with  respect to trace length $H$\label{plot:scalability-ch}}&

\includegraphics[scale=\plots]{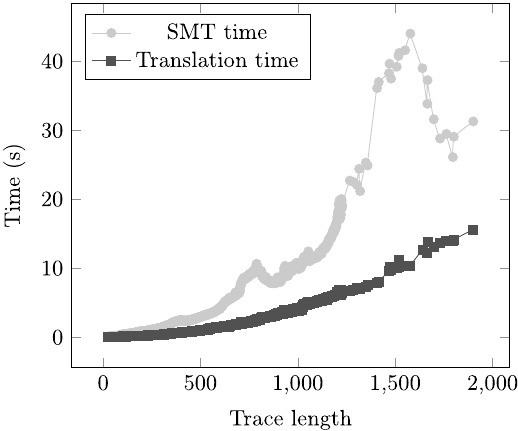}&

\includegraphics[scale=\plots]{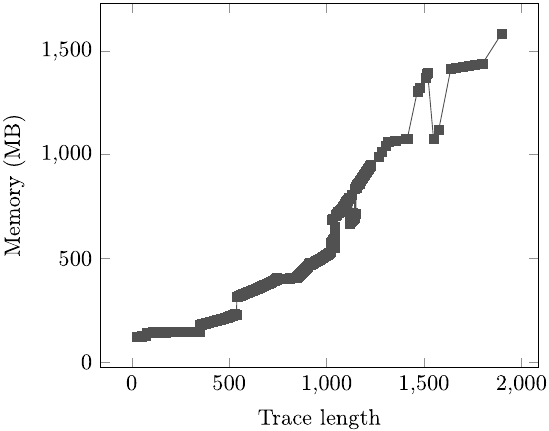}\\

\vspace{-3cm}\subcaption{Time and memory scalability of $\mathfrak{D}$ modality with respect to trace length $H$\label{plot:scalability-dh}}&

\includegraphics[scale=\plots]{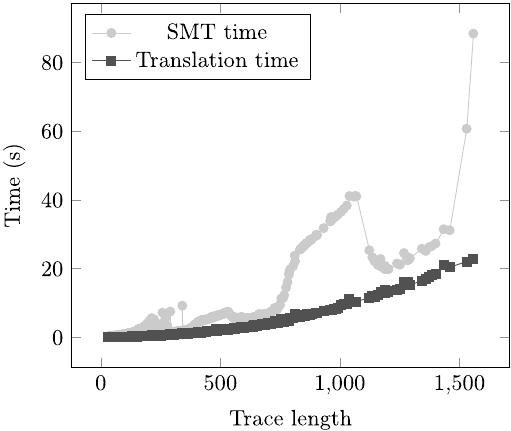}&

\includegraphics[scale=\plots]{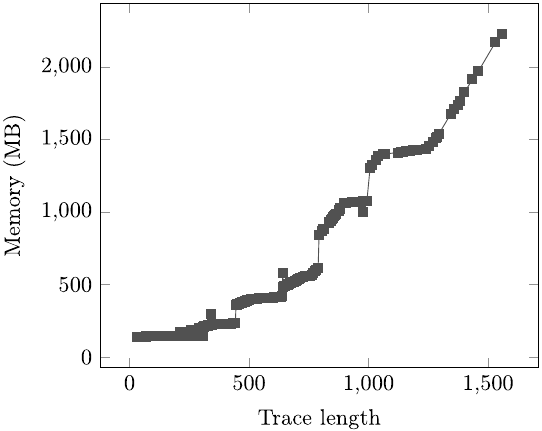}\\

\vspace{-3cm}\subcaption{Time scalability of $\mathfrak{C}$ modality
  with respect to time window $K$ (left) and the
bound $n$ (right) \label{plot:scalability-cKn}}&

\includegraphics[scale=\plots]{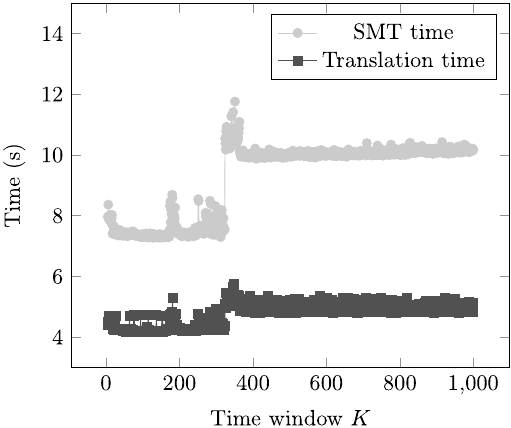}&

\includegraphics[scale=\plots]{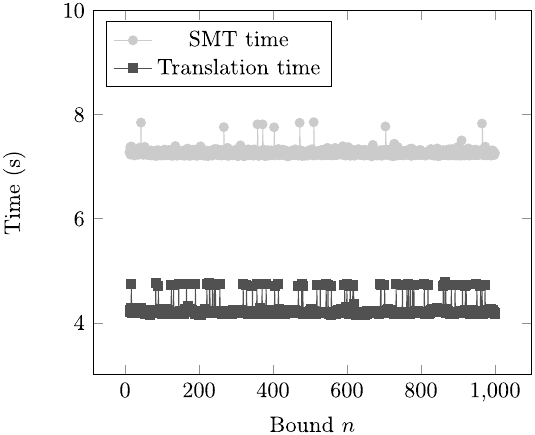}\\

\end{tabularx}
	
\caption{Evaluation of \sol-based trace checking}
\label{tab:plots}
\end{figure*}

%%% Local Variables: 
%%% mode: latex
%%% TeX-master: "soca2014"
%%% End: 

\fi

As for the scalability with respect to the other parameters, namely
the length of the time window $K$ and the bound of the comparison
operator $n$, we notice that they do not affect the resource usage,
and only introduce some non-deterministic noise in the SMT solver
time. 
\ifreport
This can be seen in Fig.~\ref{plot:scalability-cKn}, which shows the
time usage with respect to the variation of each of these two
parameters when checking formulae over a synthesized
trace of length fixed to 1000; the left plot refers to checking
formula  $\mathfrak{C}^{K}_{> 5}(p)$, while the right one refers to
checking formula  $\mathfrak{C}^{1000}_{> n}(p)$.
We omit the corresponding plots for the memory usage, since it is constant
for any value of $K$ and $n$ in the formulae.
\else
We evaluated the time and memory usage with respect to the
variation of each of these two parameters when checking properties over
a synthesized trace of length fixed to 1000.  We omit the
corresponding plots, since they show a constant value of time/memory
with respect to any value of $K$ and $n$ in the properties..
\fi

% \begin{figure*}[tb]
% \centering
% \begin{tabular}{c c}
% \input{plots/sparseness-time}&
% %\includegraphics{plots/sparseness-time-figure0.pdf}&
% \input{plots/sparseness-mem}\\
% %\includegraphics{plots/sparseness-mem-figure0.pdf}\\
% \end{tabular}

% \caption{Comparison between  \qf- and \ct-based encodings}
% \label{plots2}
% \end{figure*}

\subsection{RQ2: Comparison with the \qf-based encoding}
\label{sec:comparison-with-qf}
To investigate RQ2, we compared our approach with previous work for
trace checking of \sol~\cite{bbgks-fase2014}. 
This trace checking procedure is
based on an encoding of \sol properties into formulae of \qf, the
theory of quantifier-free integer difference logic with uninterpreted
function and predicate symbols.  
This encoding was tailored for sparse
traces, i.e., traces in which the number of time instants when events
occur is very low with respect to the length of the trace. 

The comparison focuses on how the two
approaches deal with traces of various sparseness degrees, where
sparseness is defined as the ratio between the
number of time instants in the trace where events occur and the total time
the trace spans over. 
We compared the performance of the two approaches by classifying the
generated traces into seven groups with 100\%, 50\%, 33\%, 25\%, 20\%,
16.6\%, and 14.3\% of sparseness, respectively.  We reevaluated the
approach from~\cite{bbgks-fase2014} on traces from each group and
compared time and memory requirements of both approaches. 
\ifreport
 As shown in
Fig.~\ref{plot:qfeufidl},
\else
 As shown in
the bottom plot in Fig.~\ref{plots}, 
\fi
the approach presented in this paper is more
efficient when the degree of sparseness of input traces is 25\% or
higher.  The black line in the plot shows the performance of
our approach, while the seven gray lines show our reevaluation of the
approach based on \qf, applied to traces from the seven groups
aforementioned.

\ifreport
\subsection{RQ3: Comparison with the LTL-based encoding}
\label{subsec:evaluation-encab}
To address RQ1,
we synthesized a sample
history trace of length 30 containing occurrences of an event $p$, and
considered the \sol formula $\mathfrak{C}^{K}_{> 2}(p)$, which checks whether there
have been more than two occurrences of the event $p$ within the last
$K$ time units. We varied  the length of the time window $K$
progressively from 2 to 30; the formula was always evaluated at the
last time instant of the trace. 
 Figure~\ref{plot:ltl} shows the
time and memory usage for the two translations.
The results show that the length of the formula resulting from the translation into LTL depends on 
the size of the time window $K$.  The translation 
from~\cite{bianculli13:_tale_solois} was inefficient and produced a large 
encoding that resulted in a considerable increase in time and memory
usage; our encoding addresses these issues,
paving the way for a more efficient trace checking.

\subsection{RQ4: Application to a realistic example}
\label{subsec:evaluation-running}
Here we report on the use of trace checking
to assess whether the \sol properties defined in
Sect.~\ref{sec:running} hold for the executions of the business
process described in the running example.  We performed the checks on
10  traces of length 1000 that we picked randomly from the set of
synthesized traces.  We checked the three properties as well as their
negations; the average time and memory usage, as well as their
standard deviation, are reported
in table~\ref{tab:evaluation}; these results show the practical
feasibility of our approach.

\begin{table}[t]
  \centering
\newcolumntype{A}{>{\raggedleft\arraybackslash}m{0.9cm}}
\newcolumntype{B}{>{\raggedleft\arraybackslash}m{1.1cm}}
\newcolumntype{C}{>{\centering\arraybackslash}m{0.97cm}}
\newcolumntype{M}{>{\centering\arraybackslash}m{1.15cm}}
\newcolumntype{d}[1]{D{.}{.}{#1}}

\scriptsize
\caption{Evaluation data of the running example}
\begin{tabular}{ B C d{5} d{6} d{6} d{7}}
\toprule
Property & 
Outcome & 
\multicolumn{1}{c}{ZOT} & 
\multicolumn{1}{c}{SMT} & 
\multicolumn{1}{c}{Total} & 
\multicolumn{1}{c}{Memory} \\
&
&
\multicolumn{1}c{time (s)} &
\multicolumn{1}c{time (s)} &
\multicolumn{1}c{time (s)} &
\multicolumn{1}c{(MB)} \\
\midrule
&
&
\multicolumn{4}c{(mean/standard deviation)}\\
\midrule

(QP1) &
true &
10.97/0.62&
19.70/1.55&
30.27/1.60&
931.0/0\\

(QP2) &
false &
20.77/1.37&
107.30/2.69&
125.50/2.29&
1261.7/118.24\\

(QP3) &
true &
10.56/0.46&
31.89/1.74&
42.08/1.73&
793.0/0 \\

($\neg$ QP1) &
false &
11.59/0.75&
28.00/1.67&
38.94/1.73&
932.0/0\\

($\neg$ QP2) &
true &
20.74/0.69&
324.80/159.1&
343.00/158.2&
1210.8/80.47\\

($\neg$ QP3) &
false &
11.10/0.53 &
31.37/0.55 &
41.97/0.70 &
942.0/0 \\

\bottomrule

\end{tabular}

\label{tab:evaluation}
\end{table}
\normalsize

%%% Local Variables: 
%%% mode: latex
%%% TeX-master: "soca2014"
%%% End: 

\else
\fi

%%% Local Variables: 
%%% mode: latex
%%% TeX-master: "soca2014"
%%% End: 

\section{Related Work}
\label{sec:related-work}
This work lies in the wider area of research on verification of SBAs;
we refer the reader to various
surveys~\cite{bozkurt2012:testing--verifi,salaun2010:analysis-and-ve,canfora2008:service-oriente,
  2007:test-and-analys}, illustrating approaches both for design-time
and for run-time verification of functional and QoS properties.  In
the rest of this section we focus on existing work on trace checking
and verification of quantitative properties specified in languages
similar to \sol. For a detailed discussion on \sol and related
specification languages see~\cite{bianculli13:_tale_solois}.

Finkbeiner et al.~\cite{FinkbeinerBernd2005} describe an approach to
collect statistics over run-time executions. They extend LTL to return
values from a  trace and use them to compute aggregate
properties of the trace.  However, the specification language they use
to describe the statistics to collect provides only limited support
for timing information. For example, compared to \sol, it cannot
express properties on a certain subset of an execution trace.
Furthermore, their evaluation algorithm relies on the formalism of
algebraic alternating automata. These automata are manually
built from the specification; thus making frequent changes to
the property error-prone.

In reference~\cite{Basin_etal:mfotl_aggregation} authors define an extension of metric
first-order temporal logic (MFOTL) which supports aggregation. This language 
is very similar to \sol with a general definition that supports any aggregate
operator that can be defined as a mapping from multisets to $\mathbb{Q} \cup \{\perp\}$.
The language can express aggregate properties over the values of the parameters
of relations, while \sol expresses aggregate properties on the occurrences of relations
in the temporal first-order structure.

The trace checking approach presented in~\cite{BarreBenjaminMapReduce2013}
exploits a Map-Reduce framework to validate properties of 
traces written in LTL. This work mainly focuses on recasting the trace
checking problem into a Map-Reduce framework, by distributing (sub)trace
validation tasks over many parallel sites.

In reference~\cite{Bauer:2009:FPL:1615503.1615511}, authors introduce a
specification language $PTLTL^{FO}$ (past time linear temporal logic
with first-order (guarded) quantifiers) with a counting quantifier. It
is used for expressing policies that can categorize the behavioral
patterns of a user based on its transaction history. The counting
quantifier counts the occurrences of an event from the beginning of
the trace until the position of evaluation. The difference with the
 $\mathfrak{C}$ modality of \sol is that there is no timing information:
this means one cannot specify the exact part of the  trace the
modality should consider.

In reference~\cite{Alfaroperformancereliability1997}, de Alfaro proposes pTL and
pTL* as probabilistic extensions of CTL and CTL*. These new languages
include a new modality $\mathfrak{D}$ that expresses the bound on the
average time between events.  This is achieved by using an
instrumentation clock that keeps track of the elapsed time from the
beginning of the computation until the first occurrence of a specified
event. To this end, the extended pTL formulae are evaluated on an
instrumented timed probabilistic Markov decision process.  Notice that
the $\mathfrak{D}$ modality used
in~\cite{Alfaroperformancereliability1997} differs from the one we
introduced here, since
%\sol, the $\mathfrak{D}$ modality from~\cite{Alfaroperformancereliability1997} is substantially different, since 
it computes the time passed before the first occurrence of an event, averaged
over the different computations of the underlying Markov decision
process.

%%% Local Variables: 
%%% mode: latex
%%% TeX-master: "soca2014"
%%% End: 

\section{Conclusion and Future Work}
\label{sec:concl-future-work}

The interactions among the various services participating in a
composite SBA and the provisioning of such
services can be characterized by precise specification patterns~\cite{bgps:icse2012}. The
\sol language was developed~\cite{bianculli13:_tale_solois} to express these patterns, which involve
aggregate operations on events occurring in a given time window.
In this paper, we propose an SMT-based offline trace checking procedure 
for \sol. This approach exploits a
translation of \sol into \ct, a variant of linear
temporal logic that supports counter variables.  We assess the
scalability of the approach with respect to the various parameters
involved in \sol trace checking, and we also compare it with previous work.

The use of \sol in the context of practical verification activities is
the goal of further on-going research and we intend to validate our
proposal in realistic scenarios, in collaboration with industrial
partners.  After further improvements to the translation, we also plan
to move from offline trace checking to run-time verification,
integrating ZOT and the \sol plugin into a run-time  monitoring
framework for SBAs.

%%% Local Variables: 
%%% mode: latex
%%% TeX-master: "soca2014"
%%% End: 

\ifreport
\section*{Acknowledgments}
This work has been partially supported by  the National
Research Fund, Luxembourg (FNR/P10/03).
The authors wish to thank Marcello Bersani and Matteo Pradella 
for their precious help with ZOT.
\else
\fi

%\nocite{*}
\ifreport
\bibliographystyle{plain}
\else
\bibliographystyle{IEEEtran}
\fi
\bibliography{biblio}

\end{document}